\documentclass[iop, apj]{emulateapj}
 \usepackage{graphicx}
\usepackage{natbib}
\usepackage{footnote}
\usepackage{amsmath}

\shorttitle{Introducing {\tt CAFein}, a New Computational Tool for Stellar Pulsations and Dynamic Tides}
\shortauthors{Valsecchi et al. 2012}

\begin{document}


   \title{Introducing {\tt CAFein}, a New Computational Tool for Stellar Pulsations and Dynamic Tides}


\author{F. Valsecchi, W. M. Farr, B. Willems, F.A. Rasio, and V. Kalogera}
\affil{Dept of Physics and Astronomy \& Center for Interdisciplinary Exploration and Research in Astrophysics (CIERA), Northwestern University, 2145 Sheridan Road, Evanston, IL 60208, USA.}
\email{francesca@u.northwestern.edu}


\begin{abstract}
Here we present {\tt CAFein}, a new computational tool for investigating radiative dissipation of dynamic tides in close binaries and of non-adiabatic, non-radial stellar oscillations in isolated stars in the linear regime. For the latter, {\tt CAFein} computes the non-adiabatic eigenfrequencies and eigenfunctions of detailed stellar models. The code is based on the so-called Riccati method, a numerical algorithm that has been successfully applied to a variety of stellar pulsators, and which doesn't suffer of the major drawbacks of commonly used shooting and relaxation schemes. Here we present an extension of the Riccati method to investigate dynamic tides in close binaries.
We demonstrate {\tt CAFein}'s capabilities as a stellar pulsation code both in the adiabatic and non-adiabatic regime, by reproducing previously published eigenfrequencies of a polytrope, and by successfully identifying the unstable modes of a stellar model in the $\beta$ Cephei/SPB region of the Hertzsprung-Russell diagram. Finally, we verify {\tt CAFein}'s behavior in the dynamic tides regime by investigating the effects of dynamic tides on the eigenfunctions and orbital and spin evolution of massive Main Sequence stars in eccentric binaries, and of hot Jupiter host stars. The plethora of asteroseismic data provided by the NASA's \emph{Kepler} satellite, some of which include the direct detection of tidally excited stellar oscillations, make {\tt CAFein} quite timely. Furthermore, the increasing number of observed short-period detached double white dwarfs (WD) and the observed orbital decay in the tightest of such binaries open up a new possibility of investigating WD interiors through the effects of tides on their orbital evolution.
\end{abstract}


\section{Introduction} \label{Introduction CAFein}

The current state and evolution of binary systems is affected by a wide range of physical processes, the understanding of which is important in interpreting observations. In this work we focus on dynamic tides, the tidal regime in which the free oscillation modes (eigenmodes) of one of the binary components can be 
excited by the companion's periodic tidal potential, with the driving frequency being comparable to the stellar eigenfrequencies. For this purpose, we have developed {\tt CAFein} (Code for non-Adiabatic, non-radial Forced stEllar oscillatIoNs), a novel computational tool to investigate in detail the impact of dissipation of dynamic tides in close binaries. The efficiency of non-adiabatic, dynamic tides in exchanging angular momentum between the binary orbit and the component spins in a binary depends on the tidal energy dissipation mechanism and on its strength. In addition, this new dynamic-tides tool can be used for the study of non-forced stellar pulsations. 

Thanks to NASA \emph{Kepler}'s unprecedented photometric accuracy, the effects of dynamic tides have become readily visible in electromagnetic data \citep{WelshEtAl2011, TEMBHSRCKHTIC2012}.
A phenomenal example among \emph{Kepler}'s Objects of Interest (KOI) is KOI 54, a highly eccentric binary hosting two A stars; their light curves clearly reveal that the free oscillation modes of one or both stars are tidally excited \citep{WelshEtAl2011}. The theoretical modeling of such features not only allows to further constrain the stellar and binary properties, but also potentially provides a probe to the stellar interiors, otherwise hidden to direct electromagnetic observations \citep{BurkartEtAl2012,FullerLai2012KOI54}.

Pioneering investigations targeting the observed circularization periods of massive (O, B, F) binaries identified radiative damping as the main mechanism for the dissipation of dynamic tides. In particular, \cite{Zahn1975} was the first to invoke radiative damping of tidally excitated $g$-modes in these massive Main Sequence (MS) binaries.  
A systematic comparison between the circularization periods predicted by Zahn's theory and the observations showed that many systems circularized well above the theoretically predicted period, showing evidence for a more efficient tidal dissipation mechanism \citep{GiuricinEtAl1984,NorthZahn2003,MazehTN2006}. A highly promising explanation to this discrepancy was provided by \cite{WitteSavonije2002TwoRitatingMSstars}.
The authors followed simultaneously the exchange of angular momentum between the stellar spins and the orbit due to tides and the evolution of the star's eigenfrequencies due to natural stellar evolution. This detailed analysis demonstrated that a binary can be locked into a resonant state for a prolonged period of time (the so-called \emph{resonance locking}). Such long-lasting resonances can dramatically speed up dissipation and hence tidal evolution of a binary's orbit, yielding better agreement with the observed circularization periods. Here we note that such resonance locking is unlikely to be present in solar-type stars, as it requires that the modes form standing waves. In fact, it has been shown (e.g. \citealt{GoodmanDickson1998, BarkerOgilvie2010}) that the waves resonantly excited by the tide in these stars are highly nonlinear in the core and break. They are therefore traveling waves, which suggests that the resonant
locking mechanism of Witte \& Savonije cannot operate in these systems.

Beyond the extensive work on tides in non-degenerate systems (see \citealt{Zahn2008} for a review), recent studies have focused on the effects of dynamic tides on the orbital evolution of detached binaries hosting white dwarfs (WD) and, in particular, double WDs (DWD). These binaries are widely recognized as important gravitational wave (GW) sources: they are the most numerous and guaranteed sources for the next-generation of space-based detectors sensitive to low-frequency GWs (e.g. LISA, \citealt{Danzmann1996}, and references therein, and eLISA/NGO, \citealt{NGOeLISA2012}, see also \citealt{Nelemans2001b, Nelemans2001a, NelemansEtAl2004, LiuEtAl2010b, RuiterEtAl2010} for theoretical predictions), and are currently observed electromagnetically. 
In the past few years, the Extremely Low Mass WD survey (ELM, \citealt{BrownEtAl2010,BrownEtAl2012, KilicEtAl2010, KilicEtAl2011}) successfully quintupled the number of known detached DWDs expected to merge within a Hubble time, bringing the number of systems to 24 (see Table~4 by \citealt{KilicEtAl2012} for a summary of the currently known systems) and discovering DWDs with periods down to $\simeq\,12\,$min (SDSS J065133.33+284423.3, hereafter J0651, \citealt{BrownEtAl2011,HermesEtAl2012}). The exciting J0651 system harbors a tidally deformed He WD eclipsing a C/O WD in a detached binary. Since its orbital decay was recently measured \citep{HermesEtAl2012}, J0651 is one of the cleanest astrophysical laboratories to test our understanding of tidal dissipation in these sources, and WD interiors. 

Recent work has focused on the adiabatic tidal excitation of free oscillation modes in C/O WDs. \cite{RathoreEtAl2005} used polytropic models to represent C/O WDs in eccentric binaries and focused on the $f$-mode. They found that adiabatic tides can drive the modes to high amplitudes potentially becoming nonlinear. 
Similar conclusions were reached by \cite{FullerLai2011a}, who targeted adiabatic, dynamic tides in circular binaries, using detailed C/O WD models and focusing on the excitation of $g-$modes in the close passage through a resonance. 
These authors also found that mode excitations can cause significant deviations in the orbital evolution of DWDs from the pure point-mass assumption and are very important in the spin synchronization process. Their analysis, being limited to the adiabatic regime, does not include any dissipation, which can limit the non-linear growth of the resonant modes. As a follow-up on the violation of linearity found in the adiabatic treatment, \cite{FullerLai2012Dynamic} considered the tidal excitation of gravity waves in C/O DWDs treating dissipation via the so-called ``outgoing wave boundary condition'' (BC). Such a BC implicitly assumes the waves are damped at the WD's surface via radiative damping or non-linear effects; as a result the formation of standing waves is prevented. They obtain results similar to \cite{FullerLai2011a}. Using a similar approach, \cite{FullerLai2012HeWD} extended their investigation to He WDs, focusing mainly on the effect of tidal heating and its observational signatures. 
The authors found that tidal heat is likely deposited in the outer layers of the WD and that it can dominate the WD's luminosity for the shortest orbital period binaries ($\lesssim\,15\,$min).
Recent investigations by \cite{ValsecchiFWKJ06512012} and \cite{BurkartEtAl2012HeWD} are more focused on the effect of dynamic tides on the orbital and spin evolution of DWDs. \cite{ValsecchiFWKJ06512012} applied {\tt CAFein} to a He WD model representative of the He component in J0651 to investigate the effect of linear, dissipative (non-adiabatic) dynamic tides on its orbital evolution. \cite{ValsecchiFWKJ06512012} calculated the full tidal response of the WD as multiple modes are excited simultaneously for a wide range of driving frequencies. \cite{BurkartEtAl2012HeWD} studied the effects of linear and non-linear dynamic tides in DWDs with circular orbits hosting both He and C/O WDs. In this study the dynamical tide is approximated as a superposition of standing waves and the WD response is treated as a simple harmonic oscillator with driving and dissipation. In the linear regime, the damping processes considered are thermal diffusion and turbulent convection, while in the non-linear regime the damping time for traveling waves is set by the g-mode group travel times. The radiative damping rate is also approximated in the quasi-adiabatic limit (from the star's adiabatic eigenfunctions, instead of the full non-adiabatic eigenfunctions). We refer to \cite{ValsecchiFWKJ06512012} for a discussion and comparison with the results presented by \cite{BurkartEtAl2012HeWD}.


We have developed {\tt CAFein}, a novel code to compute \emph{both} non-adiabatic, non-radial stellar oscillations in isolated stars \emph{and} forced stellar oscillations in close binaries.  
Following our understanding of tidal dissipation in non-degenerate stars, which is now able to explain the observed circularization periods in open cluster binaries \citep{Zahn1975, WitteSavonije2002TwoRitatingMSstars}, we consider radiative damping to be the main mechanism to dissipate dynamic tides in stars with radiative envelopes. 

In this paper we describe in detail the mathematical and numerical implementation of {\tt CAFein} and we present comparisons to past results found in the literature. In \S~\ref{CAFein as a Stellar Oscillations Code} we introduce {\tt CAFein} as a stellar oscillation code. In \S~\ref{The Equations Governing Non-Adiabatic and Non-Radial Stellar Pulsations} we briefly summarize the equations governing non-adiabatic and non-radial stellar oscillations and in \S~\ref{The Riccati Method} we give a detailed description of the Riccati method implemented to solve such equations. In \S~\ref{Examples of Stellar Oscillations} we test {\tt CAFein}'s results and accuracy by calculating eigenfrequencies and eigenfunctions of different stellar models. 
In \S~\ref{Testing CAFein numerical accuracy on a Polytrope: the Adiabatic Case} we calculate the adiabatic eigenfrequencies of a polytrope and check how the results change if the relevant parameters entering the Riccati method are varied. 
In \S~\ref{Our Fiducial Eigenfrequencies of a Polytrope Vs Results in the Literature} we compare these eigenferquencies with previously published results. 
In \S~\ref{Testing CAFein on a Zero Age Main Sequence Star: the Non Adiabatic Case}, we verify the reliability of {\tt CAFein} in the non-adiabatic regime by identifying the unstable modes of a stellar model in the 
$\beta$ Cephei/SPB region\footnote{SPB = slowly pulsating B-type stars} of the Hertzsprung-Russell (HR) diagram. 
In \S~\ref{Investigating Dynamic Tides With CAFein} we move onto dynamic tides and their implementation in {\tt CAFein}. 
In \S~\ref{The Tide-Generating Potential} we introduce the tide-generating potential. 
In \S~\ref{The Equations For Tidally Excited Stellar Pulsations} we summarize the equations governing tidally excited stellar pulsations and the secular evolution of the orbital elements and stellar spin. 
In \S~\ref{Extending the Riccati Method to Forced Stellar Oscillations} we describe the modifications applied to the Riccati method to solve the stellar pulsation equations when the tide-generating potential is included, and test such extension in \S~\ref{Testing the Extension of the Riccati Method to Investigate Dynamic Tides}.
In \S~\ref{Tidal Modulation of an Eigenfunction of a 5Msun MS star} we reproduce the results presented by \cite{PolflietSmeyers1990} and show that a dynamical tide can be approximated as the sum of the equilibrium tide and another part reflecting the oscillatory properties of the star itself.
In \S~\ref{Dynamic Tides Timescales in an Eccentric Binary Hosting a 5Msun MS star and a Neutron Star} we reproduce the results presented by \cite{WVHS2003} on the orbital and spin evolution timescales due to dynamic tides for an eccentric binary hosting a 5$\,M_{\odot}$ MS star and a neutron star. In \S~\ref{Dynamic Tides Timescales in an Eccentric Binary Hosting a 1.5Msun MS star and a hot Jupiter}, we compute the orbital and spin evolution timescales for a binary hosting a 1.5$M_{\odot}$ star and a hot Jupiter. We conclude in \S~\ref{CAFein: Discussion and Conclusions}. {\tt CAFein} relies on the GNU Scientific Library (GSL) both for handling the operations with matrices and for the integration of the stellar pulsation equations described below. 
\section{Computing Non-Adiabatic Stellar Pulsations with {\tt CAFein}}\label{CAFein as a Stellar Oscillations Code}
Before describing in detail {\tt CAFein}, we give a brief summary of the equations governing non-adiabatic and non-radial stellar pulsations. We refer, e.g., to \cite{UnnoEtAl1989} and \cite{GautschySaio1995} for a detailed derivation.
\subsection{The Equations Governing Non-Adiabatic and Non-Radial Stellar Pulsations}\label{The Equations Governing Non-Adiabatic and Non-Radial Stellar Pulsations}
The equations governing the non-adiabatic and non-radial stellar oscillations are the equations of mass, momentum, and energy conservation
\begin{small}
\begin{align}
&\frac{\partial\rho}{\partial t}+\nabla\cdot(\rho{\bf u}) = 0 \label{eq;massConserv}\\
&\rho\left(\frac{\partial}{\partial t}+{\bf u}\cdot\nabla\right){\bf u} = - \nabla p-\rho\nabla\Phi\label{eq;momConserv}\\
&\rho T\left(\frac{\partial}{\partial t}+{\bf u\cdot\nabla}\right)S = \rho\epsilon_{\rm N}-\nabla\cdot{\bf F}_R.\label{eq;EnergyConserv}
\end{align}
\end{small}
To these equations one must add the equations of Poisson and radiative diffusion
\begin{small}
\begin{align}
&\nabla^2\Phi = 4\pi G\rho\label{eq:Poisson}\\
&{\bf F}_R = -\frac{4ac^*}{3\kappa\rho}T^3\nabla T.\label{eq:radDiffusion}
\end{align}
\end{small}
Here $\rho$ is the density, $p$ the pressure, $T$ the temperature, ${\bf u}$ the fluid velocity, $S$ the specific entropy, $\Phi$ the gravitational potential, $\epsilon_{\rm N}$ the nuclear energy generation rate, ${\bf F}_R$ the radiative energy flux, $a$ the radiation constant, $c^*$ the speed of light, and $\kappa$ the opacity.
For simplicity, rotation, electro-magnetic and external forces, and viscosity are neglected, together with the coupling of convection and pulsations, and the perturbation of the convective flux (in the so-called ``frozen convection'' approximation). Since convection is neglected, energy transport is assumed to occur only by radiation. 
In what follows, we focus on massive stars, which have convective cores and predominantly radiative envelopes. Furthermore, the comparisons we present with earlier published results are with studies where convection is neglected as well. However, here we note that peaks in the opacity associated with Fe and He ionization lead to the formation of outer convective regions \citep{StothersChin1993}, which could affect or excite non-radial stellar pulsations \citep{CantielloLBdKSVVLY2009}.

In the star's frame we take the equilibrium state to be spherically symmetric and assume that the spatial and temporal part of the small perturbations can be written in Eulerian form as $f'({\bf r}, t) = f'(r)Y^m_l(\theta, \phi)e^{i\sigma t}$ and similarly for the Lagrangian perturbations, denoted by $\delta f$. Here $\sigma$ is an eigenfrequency (below we will denote with $\omega \equiv \sqrt{\sigma^2 R^3(GM)^{-1}}$ its dimensionless counterpart), $Y^m_l(\theta, \phi)$ is a spherical harmonic, $l$ is the harmonic degree, and $m$ the azimuthal order. 

Following the standard procedure, we then apply a small perturbation to the unperturbed star.
Perturbing and linearizing the basic Eqs.~(\ref{eq;massConserv})-(\ref{eq:radDiffusion}) and introducing 
\begin{align}
&y_{\rm 1}\,=\,\frac{\xi_{\rm r}}{r},~~y_{\rm 2}\,=\,\frac{1}{gr}\left(\frac{p'}{\rho}+ \Phi'\right)\,=\,\frac{\sigma^2 r}{g}\frac{\xi_{\rm h}}{r}\label{eq:IntroducingY1Y2}\\
&y_{\rm 3}\,=\,\frac{1}{gr}\Phi',~~y_{\rm 4}\,=\,\frac{1}{g}\frac{d\Phi'}{dr}\\
&y_{\rm 5}\,=\,\frac{\delta S}{c_{\rm p}},~~y_{\rm 6}\,=\,\frac{\delta L_{\rm R}}{L_{\rm R}}
\label{eq:IntroducingY5Y6}
\end{align}
yields 
\begin{small}
\begin{align}
r\frac{dy_{\rm 1}}{dr} = &(V_{\rm g}-3)y_{\rm 1}+\left[\frac{l(l+1)}{c_{\rm 1}\omega^2}-V_{\rm g}\right]y_{\rm 2} + V_{\rm g}y_{\rm 3}+v_{\rm t}y_{\rm 5}\label{eq:y1}\\
r\frac{dy_{\rm 2}}{dr} = &(c_{\rm 1}\omega^2 - A^*)y_{\rm 1}+(A^*-U+1)y_{\rm 2}- A^*y_{\rm 3}+v_{\rm t}y_{\rm 5}\label{eq:y2}\\
r\frac{dy_{\rm 3}}{dr} = &(1-U)y_{\rm 3}+y_{\rm 4}\label{eq:y3}\\
r\frac{dy_{\rm 4}}{dr} = &UA^*y_{\rm 1}+UV_{\rm g}y_{\rm 2}+ [l(l+1)-UV_{\rm g}]y_{\rm 3}- Uy_{\rm 4}-Uv_{\rm t}y_{\rm 5}\label{eq:y4}\\
r\frac{dy_{\rm 5}}{dr} = &V[\nabla_{\rm ad}(U - c_{\rm 1}w^2) - 4(\nabla _{\rm ad} - \nabla)+c_{\rm 2}]y_{\rm 1}+\nonumber\\
&V\left[(\nabla_{\rm ad} - \nabla)\frac{l(l+1)}{c_{\rm 1}w^2} - c_{\rm 2}\right]y_{\rm 2}+ Vc_{\rm 2}y_{\rm 3}+ \nonumber\\
&V\nabla _{\rm ad}y_{\rm 4}+V\nabla(4-k_{\rm s})y_{\rm 5}-V\nabla y_{\rm 6}\label{eq:y5}\\
r\frac{dy_{\rm 6}}{dr} = &\left[l(l+1)\frac{\nabla_{\rm ad} - \nabla}{\nabla}  - \epsilon_{\rm ad} c_{\rm 3}V\right]y_{\rm 1}+\nonumber\\
&\left[\epsilon_{\rm ad}c_{\rm 3}V + l(l+1)\left(-\frac{\nabla_{\rm ad}}{\nabla}+ \frac{c_{\rm 3}}{c_{\rm 1}w^2}\right)\right]y_{\rm 2}+ \nonumber\\
&\left[l(l+1)\frac{\nabla_{\rm ad}}{\nabla} - \epsilon_{\rm ad}c_{\rm 3}V\right]y_{\rm 3}+\nonumber\\
&\left[\epsilon_{\rm s}c_{\rm 3} - \frac{l(l+1)}{V \nabla} - i w c_{\rm 4}\right]y_{\rm 5}-\frac{d{\rm ln}L_{\rm R}}{d{\rm ln}r}y_{\rm 6}.\label{eq:y6}
\end{align}
\end{small}
Eqs.~(\ref{eq:y1})-(\ref{eq:y6}) represent the system of equations describing non-adiabatic and non-radial stellar pulsations. Here, $\boldsymbol{\xi}$ is the displacement of a fluid element from the unperturbed position ($\xi_{\rm r}$ and $\xi_{\rm h}$ are its radial and orthogonal components, respectively), $g$ is the local gravity, $\omega^2 \equiv \sigma^2 R^3(GM)^{-1}$ is the dimensionless squared eigenfrequency, $L_{\rm R}$ the radiative luminosity, and $c_{\rm 4}$ the ratio of thermal to dynamical timescale ($\tau_{\rm th}/\tau_{\rm dyn}$). We will see below that the latter is used to determine the degree of adiabaticity. The remaining terms are summarized in Table~\ref{table:stellarPulsationTerms}. Recall that the number of radial nodes in $\xi_{\rm r}$ determines the radial order $n$ of each mode. 
\begin{deluxetable}{c|c}
\tablecaption{Terms entering the equations governing non-adiabatic and non-radial stellar pulsations: $M_{\rm r}$ is the mass contained within a radius $r$, $c_{\rm p}$ is the specific heat at constant pressure.
}
\tablewidth{0pt}
\tablehead{
\colhead{Symbol} & \colhead{Expression} 
}
\startdata
$\nabla $&$\frac{d{\rm ln}T}{d{\rm ln}p}$ \\
$\nabla_{\rm ad} $&$\left( \frac{\partial{\rm ln}T}{\partial{\rm ln}p}\right)_S$\\
$\Gamma_{\rm 1} $&$\left( \frac{\partial{\rm ln}p}{\partial{\rm ln}\rho}\right)_{\rm S}$\\
$c_{\rm 1} $&$\left(\frac{r}{R}\right)^3\frac{M}{M_r}$\\
$V$&$-\frac{d{\rm ln}P}{d{\rm ln}r} = \frac{GM_r\rho}{rp}$\\
$V_{\rm g}  $&$\frac{V}{\Gamma_{\rm 1}} =\frac{gr}{c_{\rm s}^2}$\\
$U $&$\frac{d {\rm ln} M_r }{ d {\rm ln} r} =\frac{4\pi\rho r^3}{M_r}$\\
$A^* $&$rg^{-1}N^2$\\
$k_{\rm T}  $&$\left(\frac{\partial {\rm ln}k}{\partial {\rm ln}T}\right)_\rho$\\
 $k_\rho  $&$\left(\frac{\partial {\rm ln}k}{\partial {\rm ln}\rho}\right)_{\rm T}$\\
$k_{\rm ad}  $&$\left(\frac{\partial {\rm ln}k}{\partial {\rm ln}p}\right)_{\rm S} = k_{\rm T}\nabla_{\rm ad}+\frac{k_\rho}{\Gamma_{\rm 1}}$\\
$k_{S}  $&$c_{\rm p}\left(\frac{\partial {\rm ln}k}{\partial S}\right)_{\rm p} = k_{\rm T}-v_{\rm T}k_\rho$\\
$\epsilon_{\rm T}  $&$\left(\frac{\partial {\rm ln}\epsilon_{\rm N}}{\partial {\rm ln}T}\right)_\rho$\\
$\epsilon_\rho $&$ \left(\frac{\partial {\rm ln}\epsilon_{\rm N}}{\partial {\rm ln}\rho}\right)_{\rm T}$\\
$\epsilon_{\rm ad}  $&$\left(\frac{\partial {\rm ln}\epsilon_{\rm N}}{\partial {\rm ln}p}\right)_{\rm S} = \epsilon_{\rm T}\nabla_{\rm ad}+\frac{\epsilon_\rho}{\Gamma_{\rm 1}}$\\
$\epsilon_{S} $&$ c_{\rm p}\left(\frac{\partial {\rm ln}\epsilon_{\rm N}}{\partial S}\right)_{\rm p} = \epsilon_{\rm T}-v_{\rm T}\epsilon_\rho$\\
$c_{\rm 2}  $&$(k_{\rm ad} - 4\nabla_{\rm ad})V\nabla + \nabla_{\rm ad}\left(\frac{d {\rm ln} \nabla_{\rm ad}}{d{\rm ln}r}+V\right)$\\
$c_{\rm 3}  $&$\frac{4\pi r^3\rho\epsilon_{\rm N}}{L_{\rm R}}$\\
$c_{\rm 4}  $&$\frac{4\pi r^3\rho T c_{\rm p}}{L_{\rm R}}\sqrt{\frac{GM}{R^3}}$\\
$\frac{\delta k}{k}$ & $k_{\rm T}\frac{\delta T}{T}+k_{\rho}\frac{\delta \rho}{\rho} = k_{\rm ad}\frac{\delta P}{P}+k_{\rm S}\frac{\delta S}{c_{\rm p}}$\\
$\frac{\delta \epsilon_{\rm N}}{\epsilon_{\rm N}} $&$ \epsilon_{\rm T}\frac{\delta T}{T}+\epsilon_{\rho}\frac{\delta \rho}{\rho} = \epsilon_{\rm ad}\frac{\delta P}{P}+\epsilon_{\rm S}\frac{\delta S}{c_{\rm p}}$\\
$v_{\rm T}  $&$c_{\rm p}\nabla_{\rm ad}\frac{\rho T}{p}$\\
\enddata
\label{table:stellarPulsationTerms}
\end{deluxetable}

The homogeneous system of Eqs.~(\ref{eq:y1})-(\ref{eq:y6}) with the proper BCs constitute a well-posed eigenvalue problem with complex eigenvalue $\omega$. The real and imaginary part of the eigenvalue $\omega$ ($\omega_{\rm R}$ and $\omega_{\rm I}$, respectively) represent the oscillation frequency and the linear growth ($\omega_{\rm I}~<~0$) or damping ($\omega_{\rm I}~>~0$) rate, respectively.

In the star's interior, the thermal timescale is much longer than the oscillation period and the oscillation behaves almost adiabatically. Therefore, one of the inner BCs may be chosen by considering that the entropy is constant during a single oscillation ($\delta S = 0$). The other BCs at the center are given by the equations of Poisson, mass, and momentum conservation requiring that $\Phi'$, $(p'/\rho+\Phi')$, $\xi_{\rm r}$ must be regular at the center:
\begin{small}
\begin{align}
&y_{\rm 1} - \frac{ly_{\rm 2}}{c_{\rm 1}\omega^2} = 0\label{eq:BCcenter1}\\
&y_{\rm 4} - ly_{\rm 3} = 0\label{eq:BCcenter2}\\
&y_{\rm 5} = 0.\label{eq:BCcenter3}
\end{align}
\end{small}
The outer BCs are determined by considering that near the surface the Lagrangian perturbation of the pressure must vanish, by requiring the continuity of $\Phi'$ and its first derivative d$\Phi'$/d$r$, and by considering that there is no inward radiative flux.
\begin{small}
\begin{align}
&y_{\rm 1}\left\{ 1+\left[\frac{l(l+1)}{\omega^2}-4-\omega^2\right]\frac{1}{V}\right\}-y_{\rm 2}+\nonumber\\
&y_{\rm 3} \left\{ 1+\left[\frac{l(l+1)}{\omega^2}-l-1\right]\frac{1}{V}\right\} = 0\label{eq:BCsurf1}\\
&(l+1)y_{\rm 3}+y_{\rm 4} = 0\label{eq:BCsurf2}\\
&(2-4\nabla_{\rm ad}V)y_{\rm 1}+4\nabla_{\rm ad}V(y_{\rm 2}-y_{\rm 3})+4y_{\rm 5}-y_{\rm 6} = 0.\label{eq:BCsurf3}
\end{align}
\end{small}

The homogeneous system of Eqs.~(\ref{eq:y1})-(\ref{eq:y6}) greatly simplifies in the adiabatic case, since the terms involving the perturbation of the entropy ($y_{\rm 5}$) and radiative luminosity ($y_{\rm 6}$) are neglected. Furthermore, the remaining system of four Eqs.~(\ref{eq:y1})-(\ref{eq:y4}) forms an eigenvalue problem with real eigenvalue $\omega$. When computing adiabatic stellar oscillations with {\tt CAFein} we use the zero-boundary limit in which the density and pressure vanish at the stellar surface, and we substitute Eq.~(\ref{eq:BCsurf1}) with $y_{\rm 1}-y_{\rm 2}+y_{\rm 3} = 0$.
Here we note that neither the equations nor the BCs involve the azimuthal order $m$, therefore the eigenvalue is (2$l$+1)-fold degenerate with respect to $m$.

The radial distribution of the modal families inside a star is determined by the run of the Brunt-V$\ddot{\rm a}$is$\ddot{\rm a}$l$\ddot{\rm a}$ ($N$) and Lamb ($L_{\rm l}$) frequencies, as they characterize the local vibrational properties of a star.
The Lamb frequency is the inverse of the horizontal sound-crossing timescale
\begin{equation}
L_{\rm l}^2 = \frac{l(l+1)c_{\rm s}^2}{r^2}
\end{equation}
where $c_{\rm s} = \sqrt{\Gamma_{\rm 1}p/\rho}$ is the isentropic sound speed.
The Brunt-V$\ddot{\rm a}$is$\ddot{\rm a}$l$\ddot{\rm a}$ frequency is the frequency of buoyancy oscillations
\begin{equation}
N^2 = g\left(\frac{1}{\Gamma_{\rm 1}}\frac{d{\rm ln}p}{dr}-\frac{d{\rm ln}\rho}{dr}\right).
\end{equation}
In this work we follow the prescription from \cite{BrassardEtAl1991}, which accounts for the buoyancy due to the gradient in composition.
The high frequency oscillations ($\omega^2~>~L_{\rm l}^2, N^2$) have locally the characteristics of acoustic waves ($p-$modes). The low frequency oscillations ($\omega^2~<~L_{\rm l}^2, N^2$) have locally the characteristics of gravity waves ($g-$modes). 
\subsection{The Riccati Method}\label{The Riccati Method}
Here we describe the numerical method we have implemented in {\textit {CAFein}} to solve the system of Eqs.~(\ref{eq:y1})-(\ref{eq:y6}). This so-called Riccati method, as introduced by \cite{Scott1973} and extended by \cite{Davey1977}, was applied for the first time to the stellar pulsation problem by \cite{GautschyGlatzel1990}.  

Commonly, the system of differential Eqs.~(\ref{eq:y1})-(\ref{eq:y6}) is solved using relaxation and shooting schemes. The Riccati method differs from such techniques mainly in the type of equations that have to be solved; from a technical point of view it is really a shooting method.
According to the Riccati method the linear first-order ordinary differential system describing a boundary eigenvalue problem is transformed into a numerically stable, non-linear initial value problem. This initial-value problem is then solved using a shooting method, where the eigenfrequency is the only shooting parameter to be iterated. 

For a linear two-point boundary value problem in which the solutions change very rapidly, like the problem describing stellar oscillations, the advantages of the Riccati method become most clear. As pointed out by \cite{TakataLoffler2004}, if commonly-used shooting methods are adopted, it is difficult to satisfy the matching condition in a numerically stable manner, as the eigenfunctions at the fitting point are strongly dependent on their values at the star's boundaries. On the other hand, if Henyey-type relaxation methods \citep{HFG1964} are used, the accuracy of the eigenfunctions decreases where their absolute values are very small. If small amplitude eigenfunctions can not be resolved, node counting is affected and this, in turn, affects the correct determination of the radial order of the mode.

Even though the Riccati method has been proven to be much more stable than the techniques described above, its higher numerical stability comes at the expenses of potentially higher computational times and a less straightforward implementation. Nonetheless, this method has been extensively and successfully applied to a variety of stellar and WD pulsators \citep{GautschyGlatzel1990,GautschyGlatzel1990b,GautschyGlatzel1991, GlatzelGautschy1992, GlatzelKiriakidis1993, GautschyLoeffler1996, GLF1996, SchenkerGautschy1998, Loffler2000}.

Here we provide a detailed explanation of the Riccati method, which we have implemented closely following \cite{GautschyGlatzel1990} and \cite{TakataLoffler2004}.

In what follows the subscripts ``R'' and ``I'' denote the real and imaginary part of complex quantities, respectively.
\subsubsection{The Riccati Equation}\label{The Riccati Equation}
We start by writing the original system of Eqs.~(\ref{eq:y1})-(\ref{eq:y6}) in the form
\begin{equation}
\frac{d{\bf y}}{dr} = {\bf M}{\bf y}=
\left(
\begin{array}{cc}
 {\bf A} & {\bf B} \\
 {\bf C} & {\bf D} \\
\end{array}
\right){\bf y}
\label{eq:dyABCDy}
\end{equation}
If the number of elements in the vector {\bf y} is $N$, then ${\bf M}$ is a square matrix of size $N\times N$, while ${\bf A, B, C}$, and ${\bf D}$ are square matrices of size $J$, with $N~=~2J$.
Next, we introduce two vectors ${\bf u_{\rm Ric}}$ and ${\bf v_{\rm Ric}}$ of size $J$ which store the first and last $J$ components of {\bf y}, respectively:
\begin{equation}
{\bf y} = \left(
\begin{array}{c}
 {\bf u_{\rm Ric}} \\
 {\bf v_{\rm Ric}} \\
\end{array}
\right).
\label{eq:yASuv}
\end{equation}
Then, equation (\ref{eq:dyABCDy}) can be rewritten as:
\begin{equation}
\frac{d}{dr} \left(
\begin{array}{c}
 {\bf u_{\rm Ric}} \\
 {\bf v_{\rm Ric}} \\
\end{array}
\right) = 
\left(
\begin{array}{cc}
 {\bf A} & {\bf B} \\
 {\bf C} & {\bf D} \\
\end{array}
\right)\left(
\begin{array}{c}
 {\bf u_{\rm Ric}} \\
 {\bf v_{\rm Ric}} \\
\end{array}
\right).
\label{eq:duvABCDuv}
\end{equation}
From Eq.~(\ref{eq:duvABCDuv}) we can derive two separate equations for ${\bf u_{\rm Ric}}$ and ${\bf v_{\rm Ric}}$
\begin{align}
&\frac{d{\bf u_{\rm Ric}}}{dr} = {\bf Au_{\rm Ric} +  Bv_{\rm Ric}}\label{eq:duAuBv}\\
&\frac{d{\bf v_{\rm Ric}}}{dr} = {\bf Cu_{\rm Ric} +  Dv_{\rm Ric}}\label{eq:dvCuDv}
\end{align}
Defining the Riccati matrix ${\bf R}$ as
\begin{equation}
{\bf u_{\rm Ric} = R v_{\rm Ric}}
\label{eq:uRv}
\end{equation}
from Eqs.~(\ref{eq:duAuBv}) and (\ref{eq:dvCuDv}) it is straightforward to show that $R$ satisfies
\begin{equation}
\frac{d{\bf R}}{dr} = {\bf B+AR-RD-RCR.}
\label{eq:dR}
\end{equation}
Equation~(\ref{eq:dR}) denotes the new system of non-homogeneous and non-linear differential equations that will be integrated instead of the original stellar pulsation problem. Equation~(\ref{eq:dR}) has to be solved together with $J$ homogeneous BCs at both extrema of the integration interval.
The most general form of the BCs can be written as 
\begin{equation}
{\bf Pu_{\rm Ric} = Qv_{\rm Ric}}
\label{eq:BC_EF}
\end{equation}
where {\bf P} and {\bf Q} are $J\times J$ matrices. As either ${\bf u_{\rm Ric}}$ or ${\bf v_{\rm Ric}}$ can be considered as arbitrary, Eq.~(\ref{eq:BC_EF}) uniquely determines {\bf R}.
Given the form of the non-adiabatic BCs at the star's center (\ref{eq:BCcenter1})-(\ref{eq:BCcenter3}) and surface (\ref{eq:BCsurf1})-(\ref{eq:BCsurf3}), we chose the vectors ${\bf u_{\rm Ric}}$ and ${\bf v_{\rm Ric}}$ as 
\begin{equation}
{\bf u_{\rm Ric}} = \left(
\begin{array}{c}
 {y_{\rm 1}} \\
 {y_{\rm 4}} \\
 {y_{\rm 5}} \\
\end{array}
\right),~~ 
{\bf v_{\rm Ric}} = \left(
\begin{array}{c}
 {y_{\rm 2}} \\
 {y_{\rm 3}} \\
 {y_{\rm 6}} \\
\end{array}
\right).
\label{eq:initialUV}
\end{equation}
The non-adiabatic initial conditions on the Riccati matrix ${\bf R}$ at the star's center and surface thus become
\begin{align}
&{\bf R_{\rm c}} = \left(
\begin{array}{ccc}
 \frac{l}{c_{\rm 1}\omega^2} & 0 & 0\\
0 & l & 0\\
0 & 0 & 0\\
\end{array}
\right)\label{eq:initialRcenter}\\
&{\bf R_{\rm s}} = \left(
\begin{array}{ccc}
 \frac{V}{V+z_{\rm 1}} &  -\frac{V+z_{\rm 2}}{V+z_{\rm 1}} & 0\\
0 &  -(l+1) & 0\\
 \frac{V(1+2\nabla_{\rm ad}z_{\rm 1})}{2(V+z_{\rm 1})} &  \frac{V+z_{\rm 2}-2\nabla{\rm ad}V(z_{\rm 2}-z_{\rm 1})}{2(V+z_{\rm 1})} & \frac{1}{4}\\
\end{array}
\right)\label{eq:initialRsurface}
\end{align}
where the subscripts ``c'' and ``s'' denote the center and the surface, respectively, and for ease of notation we have introduced $z_{\rm 1} = l(l+1)/\omega^2 - 4 - \omega^2$ and $z_{\rm 2} = l(l+1)/\omega^2 - l - 1$. In the adiabatic case, the BCs on the Riccati matrix reduce to
\begin{align}
&{\bf R_{\rm c}} = \left(
\begin{array}{cc}
 \frac{l}{c_{\rm 1}\omega^2} & 0 \\
0 & l \\
\end{array}
\right)\label{eq:initialRcenterAdiabatic}\\
&{\bf R_{\rm s}} = \left(
\begin{array}{cc}
1 &  -1\\
0 &  -(l+1)\\
\end{array}
\right).
\label{eq:initialRsurfaceAdiabatic}
\end{align}
\subsubsection{The Calculation of the Eigenfrequencies}\label{The Calculation of the Eigenfrequencies}
For a given frequency $\omega$, Eq.~(\ref{eq:dR}) is integrated twice: a first integration is performed outward from the star's center to a conveniently chosen fitting point $r_{\rm fit}$ with initial conditions~(\ref{eq:initialRcenter}) yielding matrix ${\bf R}^{\rm out}$($r$). A second integration is performed inward from the star's surface to $r_{\rm fit}$ with initial conditions~(\ref{eq:initialRsurface}) yielding matrix ${\bf R}^{\rm in}$($r$). The frequency $\omega$ is an eigenfrequency, when the eigenfunctions $y_{i}$ ($i$~=~1~$\rightarrow$~6) are continuous at the fitting point:
\begin{align} 
&{\bf u_{\rm Ric}}^{\rm in}(r_{\rm fit}) = {\bf u_{\rm Ric}}^{\rm out}(r_{\rm fit})\label{eq:continuityUV1}\\
&{\bf v_{\rm Ric}}^{\rm in}(r_{\rm fit}) = {\bf v_{\rm Ric}}^{\rm out}(r_{\rm fit}).
\label{eq:continuityUV}
\end{align} 
Eqs.~(\ref{eq:continuityUV1}) and (\ref{eq:continuityUV}) are equivalent to
\begin{equation} 
[{\bf R}^{\rm in}(r_{\rm fit}) - {\bf R}^{\rm out}(r_{\rm fit})]{\bf v_{\rm Ric}} = 0.
\label{eq:riccatiConditionAtFittingV}
\end{equation} 
A necessary condition for Eq.~(\ref{eq:riccatiConditionAtFittingV}) to have a non trivial solution yields 
\begin{equation} 
{\rm det}[{\bf R}^{\rm in}(r_{\rm fit}) - {\bf R}^{\rm out}(r_{\rm fit})] = 0.
\label{eq:riccatiConditionAtFitting}
\end{equation} 
Since $\omega$ is complex, we first scan the parameter space in $\omega_{\rm R}$ setting $\omega_{\rm I} = 0$. At this stage we find the interval in $\omega_{\rm R}$ across which the real part of expression~(\ref{eq:riccatiConditionAtFitting}) crosses zero. Next, the values of $\omega_{\rm R}$ at the extrema of this interval are taken together with the values of ${\rm det}[{\bf R}^{\rm in}(r_{\rm fit}) - {\bf R}^{\rm out}(r_{\rm fit})]$ as initial guesses for the iteration of the exact eigenfrequencies. During this iteration we use a complex secant method to find the exact values of $\omega_{\rm R}$ and $\omega_{\rm I}$ for which both the real and imaginary part of condition~(\ref{eq:riccatiConditionAtFitting}) are satisfied.
Following \cite{TakataLoffler2004} we chose the fitting point $r_{\rm fit}$ based on the behavior of the Brunt-V$\ddot{\rm a}$is$\ddot{\rm a}$l$\ddot{\rm a}$ and Lamb frequencies. Specifically, for a given frequency $\omega_{\rm R}$ we pick the innermost point where $N^{2}$ or $L_l^{2}$ $\simeq \omega_{\rm R}^{2}$. For low frequency $g$-modes or high frequency $p$-modes, the fitting point where $N^2\simeq \omega_{\rm R}^{2}$ or $L^2 \simeq \omega_{\rm R}^{2}$, respectively, is very close to the star's center.
Scanning the parameter space in $\omega$ following the procedure described above yields all eigenfrequencies.
\subsubsection{Reimbedding}\label{Reimbedding}
During the calculation of the eigenfrequencies (\S~\ref{The Calculation of the Eigenfrequencies}), ${\bf v_{\rm Ric}}$ might vanish and therefore ${\bf R}$ becomes singular (recall Eq.~(\ref{eq:uRv})). In this case, we avoid the singularity by re-defining ${\bf u_{\rm Ric}}$ and ${\bf v_{\rm Ric}}$ according to the so-called ``reimbedding'' procedure, as described by \cite{TakataLoffler2004}.
The new ${\bf u_{\rm Ric}'}$ and ${\bf v_{\rm Ric}'}$ are a transformation of the original ${\bf u_{\rm Ric}}$ and ${\bf v_{\rm Ric}}$  and are given by
\begin{equation}
\left(
\begin{array}{c}
 {\bf u_{\rm Ric}'} \\
 {\bf v_{\rm Ric}'} \\
\end{array}
\right) \equiv
{\bf T}\left(
\begin{array}{c}
 {\bf u_{\rm Ric}} \\
 {\bf v_{\rm Ric}} \\
\end{array}
\right)
=\left(
\begin{array}{cc}
 {\bf T_{\rm 00}} & {\bf T_{\rm 01}} \\
 {\bf T_{\rm 10}} & {\bf T_{\rm 11}} \\
\end{array}
\right)\left(
\begin{array}{c}
 {\bf u_{\rm Ric}} \\
 {\bf v_{\rm Ric}} \\
\end{array}
\right)
\label{eq:permuteUV}
\end{equation}
where ${\bf T}$ is a square $N\times N$ matrix, while ${\bf T_{\rm ij}}$ are square $J\times J$ submatrices (recall $N = 2J$). For the transformed Riccati matrix ${\bf R'}$ we can write
\begin{equation} 
{\bf R' = ( T_{\rm 00}R+ T_{\rm 01})(T_{\rm 10}R+ T_{\rm 11}})^{-1},
\label{eq:Rpermuted}
\end{equation} 
choosing matrix  ${\bf T}$ such that
\begin{equation} 
{\rm det}({\bf T_{\rm 10}R+ T_{\rm 11}})\neq 0.
\end{equation} 
The transformed Riccati matrix ${\bf R'}$ is still a solution for Eq.~(\ref{eq:dR}), provided that ${\bf A, B, C,}$ and ${\bf D}$ are replaced by their prime (') counterparts.
\begin{equation}
\left(
\begin{array}{cc}
 {\bf A'} & {\bf B'} \\
 {\bf C'} & {\bf D'} \\
\end{array}
\right) \equiv
T\left(
\begin{array}{cc}
 {\bf A} & {\bf B} \\
 {\bf C} & {\bf D} \\
\end{array}
\right)T^{-1}
\label{eq:ABCDprime}
\end{equation}
 During the calculation of the eigenfrequencies, we apply this procedure whenever the Eucledian norm of ${\bf R}$ ($||{\bf R}|| = \sqrt{\sum_{i=0}^J\sum_{j=0}^JR_{\rm ij}^{2}}$) goes above a certain pre-defined value, switching to the permuted ${\bf R'}$ with the lowest $||{\bf R'}||$ (see \S~\ref{The Calculation of the Eigenfunctions} for the permutation criterion adopted during the calculation of the eigenfunctions). Here we note that for the fully non-adiabatic case, the integration variables $y_{i}$ ($i$~=~1~$\rightarrow$~6) are complex. In this configuration, ${\bf u_{\rm Ric}}$ and ${\bf v_{\rm Ric}}$ have size 6, and we chose them so that 
\begin{align}
&{\bf u_{\rm Ric}} = \left(
\begin{array}{c}
 {y_{\rm 1, R}} \\
 {y_{\rm 4, R}} \\
 {y_{\rm 5, R}} \\
 {y_{\rm 1, I}} \\
 {y_{\rm 4, I}} \\
 {y_{\rm 5, I}} \\
\end{array}
\right) = 
\left(
\begin{array}{c}
 {{\bf u}_{\rm Ric, R}} \\
 {{\bf u}_{\rm Ric, I}} \\
\end{array}
\right) \nonumber\\ 
&{\bf v_{\rm Ric}} = \left(
\begin{array}{c}
 {y_{\rm 2, R}} \\
 {y_{\rm 3, R}} \\
 {y_{\rm 6, R}} \\
 {y_{\rm 2, I}} \\
 {y_{\rm 3, I}} \\
 {y_{\rm 6, I}} \\
\end{array}
\right) = 
\left(
\begin{array}{c}
 {{\bf v}_{\rm Ric, R}} \\
 {{\bf v}_{\rm Ric, I}} \\
\end{array}
\right).
\label{eq:initialUVcomplex}
\end{align}
The resulting ${\bf R}$ has a size 6$\,\times\,$6 and  the search for the permutation yielding ${\bf R'}$ with the minimum norm would result in long computational times. However, given that
\begin{equation}
 \left(
\begin{array}{c}
 {{\bf u}_{\rm Ric, R}} \\
 {{\bf u}_{\rm Ric, I}} \\
\end{array}
\right)
 = \left(
\begin{array}{cc}
 {\bf R}_{\rm 1} & {\bf R}_{\rm 2} \\
 {\bf R}_{\rm 3} & {\bf R}_{\rm 4} \\
\end{array}
\right)  \left(
\begin{array}{c}
 {{\bf V}_{\rm Ric, R}} \\
 {{\bf V}_{\rm Ric, I}} \\
\end{array}
\right)
\label{eq:Rcomplex}
\end{equation}
we minimize the numbers of trials finalized to find the permutation yielding ${\bf R'}$ with the minimum norm by only permuting the (dominant) real part of ${\bf R}$ (${\bf R}_{\rm 1}$ and ${\bf R}_{\rm 4}$) and applying that same permutation to the imaginary part (${\bf R}_{\rm 2}$ and ${\bf R}_{\rm 3}$). 
In the adiabatic case, instead, only Eqs.~(\ref{eq:y1})-(\ref{eq:y4}) are solved (the terms containing $y_{\rm 5}$ and $y_{\rm 6}$ are neglected) and the eigenfrequency is purely real. This reduces the size of the Riccati matrix to 2$\times$2 and we scan on all possible permutations to minimize $||{\bf R'}||$.
\subsubsection{The Calculation of the Eigenfunctions}\label{The Calculation of the Eigenfunctions}
Once the eigenfrequencies of interest have been determined, the calculation of the eigenfunctions for a particular eigenfrequency proceeds as follows. Eq.~(\ref{eq:dR}) is integrated as described in \S~\ref{The Calculation of the Eigenfrequencies} and the components of ${\bf R}$ ($R_{\rm ij}$) are stored together with the permutations applied. If the adiabatic eigenfunctions are computed, we permute the Riccati matrix as described in \S~\ref{Reimbedding}. In the non-adiabatic case, we track the behavior of the Eucledian norm of {\bf R} during the integration and apply a permutation every time $|| R ||$ has a maximum.
Once $R_{\rm ij}$ have been evaluated both for ${\bf R}^{\rm in}(r)$ and ${\bf R}^{\rm out}(r)$, the eigenfunctions are readily calculated by solving
\begin{equation}
\frac{d{\bf v_{\rm Ric}}}{dr} = {\bf (CR+D)v_{\rm Ric}}
\label{eq:dvEigenfunc}
\end{equation}
together with Eq.~(\ref{eq:uRv}). It is straightforward to derive Eq.~(\ref{eq:dvEigenfunc}) from Eqs.~(\ref{eq:dvCuDv}) and (\ref{eq:uRv}). Eq.~(\ref{eq:dvEigenfunc}) is integrated twice, from the fitting point $r_{\rm fit}$ to the star's center and from $r_{\rm fit}$ to the surface. 
At this stage, we do not integrate ${\bf R}$ again, but we interpolate the already calculated $R_{\rm ij}$ using linear interpolation in the adiabatic regime and a third order polynomial \citep{Steffen1990} in the non-adiabatic regime, and apply the same permutations used during their calculation.
The initial value of ${\bf v_{\rm Ric}}$ is given by a non-trivial solution of Eq.~(\ref{eq:riccatiConditionAtFittingV}).
Clearly, during the integration from $r_{\rm fit}$ to the center ($r_{\rm fit}$ to the surface) ${\bf R}^{\rm out}$ (${\bf R}^{\rm in}$) must be used.
This numerical scheme where ${\bf v_{\rm Ric}}$ and ${\bf R}$ are ${\textit{not}}$ integrated together and are integrated in the opposite directions is necessary for numerical stability \citep{Sloan1977}.
\begin{deluxetable}{lr}
\tablecolumns{2}
\tabletypesize{\scriptsize}
\tablewidth{0pc}
\tablecaption{Units of Physical Quantities. The total mass and radius are denoted with $M$ and $R$, respectively.}
\tablehead{
\colhead{Unit of:} & \colhead{Unit}}
\startdata
Length & $R$ \\
Mass & $M$ \\
Time & $\sqrt{R^{3} /(GM)}$\\
Temperature & $GM^{2}/(R R_{\rm gas})$\\
Energy & $GM^{2}R^{-1}$
\enddata
\label{Tab:unitsOfPhysicalQuantities}
\end{deluxetable}
\subsection{Examples of Free Stellar Oscillations and Tests}\label{Examples of Stellar Oscillations}
In this section, we present a series of tests for {\tt CAFein} both internal and against published results in the literature. For the latter, we compute the eigenfrequencies of a polytropic model and compare them with results in the literature which rely both on the Riccati method (\citealt{TakataLoffler2004}, hereafter TL04) and on standard relaxation and shooting techniques (\citealt{CDM1994}, hereafter CDM94).  
In what follows, we refer to the eigenfrequency of the purely adiabatic problem with $\omega_{\rm Ad}$, while we denote with $\omega$ the non-adiabatic eigenfrequency. The subscripts ``R'' and ``I'' have their usual meaning.
\subsubsection{Testing {\tt CAFein}'s numerical accuracy on a Polytrope: the Adiabatic Case}\label{Testing CAFein numerical accuracy on a Polytrope: the Adiabatic Case}
\begin{figure}
\epsscale{0.9}
\plotone{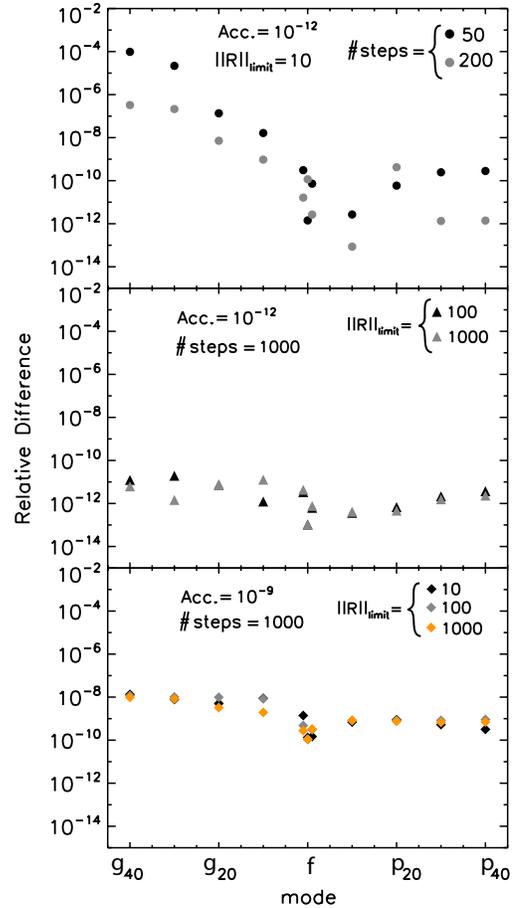}
\caption{Test on {\tt CAFein}'s performances if the required integration accuracy (both relative and absolute), $||{\bf R}||_{{\rm limit}}$, and resolution adopted during the scan of the parameter space in $\omega$ are varied. For the latter, we consider a randomly chosen interval [$\omega_{\rm Ad}$-0.1, $\omega_{\rm Ad}$+0.1] across each eigenfrequency. Here we set $l\,=\,$2, but the same test performed on the $l\,=\,3$ eigenfrequencies yielded similar results. The relative difference is the absolute value of the difference between our fiducial eigenfrequencies and the ones calculated by changing the parameters mentioned above, divided by our fiducial values. {\emph{Top: }} the integrator accuracy and $||{\bf R}||_{{\rm limit}}$ are kept fixed, while the eigenfrequency scan resolution (\# steps) is varied; {\emph{Middle}: } the integrator accuracy and eigenfrequency scan resolution are kept fixed, while  $||{\bf R}||_{{\rm limit}}$ is varied; {\emph{Bottom: }} same as the middle panel, but for a lower integrator accuracy (note that an accuracy of $10^{-12}$ and $||{\bf R}||_{{\rm limit}}\,=\,10$ determine our fiducial values). Some of the data overlap and are not visible.}
\label{fig:StabilityCAFein_withPolytrope} 
\end{figure}

In this section, we apply {\tt CAFein} to a polytropic model to compute its eigenfrequencies. We then test the numerical accuracy of our calculation by changing some of the relevant parameters entering the computation of the Riccati matrix and analyzing how the eigenfrequencies vary. In what follows, when we mention the integrator accuracy adopted in our calculation, we refer both to the absolute and relative accuracies.

We create a polytrope with index $n$~=~3, assuming an ideal gas with $\Gamma_{\rm 1}$~=~5/3. Below, we indicate the polytropic index with $n_{\rm p}$ to avoid confusion with the index denoting the radial order of a mode.
We first solve the Lane-Emden equation using a variable step 4th order Runge-Kutta integrator with an accuracy requirement of $10^{-12}$. We then iterate the integration until we reach a resolution between two consecutive mesh points of $\Delta r \leq 5\times 10^{-5}$ (here we are following TL04, but adopting a slightly higher resolution). 
We conveniently pass onto dimensionless quantities by expressing the physical quantities in the units listed in Table~\ref{Tab:unitsOfPhysicalQuantities}. This table also lists the units that will be used in the non-adiabatic regime.
During the calculation of the adiabatic eigenfrequencies, we use again a variable step 4th order Runge-Kutta integrator and we interpolate the various polytropic parameters using a third order polynomial \citep{Steffen1990}.
As an example, we report some of the calculated eigenfrequencies $\omega_{\rm Ad}$ in Table~\ref{Tab:polytrope_n3_EigenfreqAdiabatic}. Here we set the required integration accuracy to $10^{-12}$, the limit on the Eucledian norm of ${\bf R}$ ($||{\bf R}||_{{\rm limit}}$) used for reimbedding (\S~\ref{Reimbedding}) to 10, and perfomed 1000 integrations in $\omega$ on a randomly chosen interval [$\omega_{\rm Ad}$-0.1, $\omega_{\rm Ad}$+0.1] across each eigenfrequency. In what follows, we will refer to the eigenfrequencies listed in Table~\ref{Tab:polytrope_n3_EigenfreqAdiabatic} as our ``fiducial'' ones.

Next, we test the stability of {\tt CAFein} with respect to the parameters mentioned above. We re-compute the same eigenfrequencies for different integrator's accuracies ($10^{-9}$ and $10^{-12}$), $||{\bf R}||_{{\rm limit}}$ (10, 100, and 1000), and number of integrations across each eigenfrequency (50, 200, and 1000), fixing the width of each integration interval as described above, and compare them with our fiducial values.  The outcome of this test is summarized in Fig.~\ref{fig:StabilityCAFein_withPolytrope}, where it is clear that our results are sensitively affected only by the resolution adopted during the scan of the parameter space in $\omega$ (i.e. the eigenfrequency scan resolution, see top plot). Varying $||{\bf R}||_{{\rm limit}}$ while keeping the number of iterations and integrator accuracy fixed, the relative difference with our fiducial eigenfrequencies is between $10^{-13}-10^{-11}$ (middle plot in Fig.~\ref{fig:StabilityCAFein_withPolytrope}). Decreasing the integrator accuracy by 3 order of magnitudes, the relative difference with our fiducial eigenfrequencies is $\lesssim~10^{-8}$ (bottom plot in Fig.~\ref{fig:StabilityCAFein_withPolytrope}). 
The same test performed on the $l\,=\,$3 eigenfrequencies yielded similar results. 

Once the eigenfrequencies are known, the eigenfunctions can be readily calculated as described in \S~\ref{The Calculation of the Eigenfunctions}. 
The eigenfunctions should be orthogonal. In fact, as \cite{FullerLai2011a} point out, the numerical determination of an eigenfunction might be contaminated by other eigenfunctions
\begin{equation}
(\boldsymbol{\xi}_{\rm \alpha})_{\rm num}=h_{\rm \alpha} \boldsymbol{\xi}_{\rm \alpha}+h_{\rm 0} \boldsymbol{\xi}_{\rm 0}+h_{\rm 1} \boldsymbol{\xi}_{\rm 1}+...
\label{eq:Xi_numerical}
\end{equation}
where the subscript $\alpha$ denotes the order of the mode and the displacement from the equilibrium position $\boldsymbol{\xi}_{\rm \alpha}$ for the mode $\alpha$ can be written as
\begin{equation}
\boldsymbol{\xi}_{\rm \alpha} = [\xi_{\rm r,\alpha}(r){\bf e}_{\rm r} + \xi_{\rm h,\alpha}(r){\bf e}_{\rm h} \nabla]Y^{\rm m}_{\rm l}(\theta,\phi).
\end{equation}
The terms with subscript ``0'' in Eq.~(\ref{eq:Xi_numerical}) refer to the $f$-mode and the various  coefficients are given by
\begin{equation}
h_{\rm i} = <\boldsymbol{\xi}_{\rm i}|\boldsymbol{\xi}_{\rm \alpha}> = \int_0^1 r^2 \rho [\xi_{\rm r,i}\xi_{\rm r,\alpha}+l(l+1)\xi_{\rm h,i}\xi_{\rm h,\alpha}] \mathrm{d} r
\label{eq:def_h_i}
\end{equation}
where the various eigenfunctions in Eq.~(\ref{eq:def_h_i}) are the ones calculated numerically. 
To derive Eq.~(\ref{eq:def_h_i}), we used the normalization of the spherical harmonics as given by \cite{UnnoEtAl1989}:
\begin{equation}
\int_0^{2\pi}\int_0^\pi Y^m_l(\theta, \phi) Y^{m'}_{l'}(\theta, \phi)~{\rm sin}\theta~{\rm d}\theta ~{\rm d}\phi = \delta_{ll'}\delta_{mm'}
\end{equation}
where $\delta_{ll'}$ and $\delta_{mm'}$ are the Kronecker deltas. 
Since the $f-$mode gives the dominant contribution in Eq.~(\ref{eq:Xi_numerical}), we normalize the eigenfunctions so that $h_{\rm \alpha} = <\boldsymbol{\xi}_{\rm \alpha}| \boldsymbol{\xi}_{\rm \alpha}>\,=\,1$ and we take $(\boldsymbol{\xi}_{\rm \alpha})_{num}$ to accurately represent the actual $\boldsymbol{\xi}_{\rm \alpha}$ if $|h_{\rm 0}|  \ll 1$.
The values of the coefficient $h_{\rm 0}$ for the $f-$mode and the first five $p-$ and $g-$modes of harmonic degree $l$~=~2 for the polytropic model considered are reported in Table~\ref{Tab:polytrope_n3_l2_orthogonality}. The results show that orthogonality is satisfied to the expected accuracy of our eigenfunctions, $\simeq\sqrt{10^{-12}}$.

Recall that {\tt CAFein} has been developed to investigate dynamic tides in close binaries. Our focus on the harmonic degree $l\,=\,2$ will become clear in \S~\ref{The Tide-Generating Potential}, where we introduce the tide-generating potential. 
\subsubsection{Comparing our Fiducial Polytrope Eigenfrequencies to Polytrope Results in the Literature }\label{Our Fiducial Eigenfrequencies of a Polytrope Vs Results in the Literature}
\begin{figure}
\epsscale{0.9}
\plotone{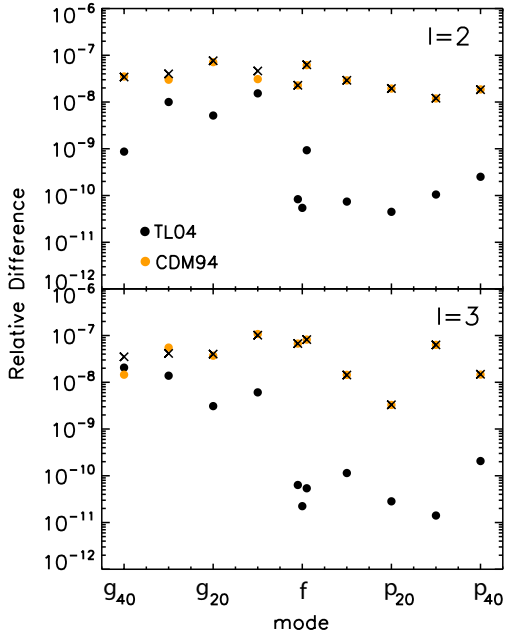}
\caption{Relative difference between our fiducial eigenfrequencies and the one presented by TL04 and CDM94. As in Table~\ref{Tab:polytrope_n3_EigenfreqAdiabatic}, we consider a polytropic model with $n_{\rm p}\,=\,3$. \emph{Top: } $l\,=\,2$; \emph{Bottom: } $l\,=\,3$. The relative difference is the absolute value of the difference between our fiducial eigenfrequencies and the ones reported by TL04 or CDM94, divided by the fiducial values. For comparison, we denote with ``$\times$'' the relative difference between TL04 and CDM94.}
\label{fig:goldenComparisonWithPreviousResults} 
\end{figure}

\begin{deluxetable}{c|c|c|c}[!h]
\tablecolumns{4}
\tabletypesize{\scriptsize}
\tablewidth{0pc}
\tablecaption{Eigenfrequencies for a polytropic model with $n_{\rm p}\,=\,3$. Here we used an accuracy requirement for the integrator of $10^{-12}$, $||{\bf R}||_{{\rm limit}}\,=\,$10, and 1000 integrations in $\omega$ on a randomly chosen interval [$\omega_{\rm Ad}$-0.1, $\omega_{\rm Ad}$+0.1] across each eigenfrequency. 
See Fig.~\ref{fig:StabilityCAFein_withPolytrope} for a test on the numerical accuracy of the eigenfrequencies calculated in this work and  Fig.~\ref{fig:goldenComparisonWithPreviousResults} for the relative differences between our results  and the ones presented by TL04  and CDM94. For TL04 the values are taken from Table~1 of their paper, while for CDM94 we used Table~2 and 4 of their paper.}
\tablehead{
\colhead{Ref.} & \colhead{mode} & \colhead{$\omega_{\rm Ad}^2, l\,=\,2 $} & \colhead{$\omega_{\rm Ad}^2, l\,=\,3 $}}
\startdata
This Work &&  2.7777508707$\times 10^{3}$ &  2.8357237896$\times 10^{3}$\\
TL04 &          $p_{\rm 40}$ &  2.7777508750$\times 10^{3}$ &  2.8357239610$\times 10^{3}$\\
CDM94 &&  2.7777509770$\times 10^{3}$ &  2.8357238770$\times 10^{3}$\\
\hline
This Work &&  1.6220938771$\times 10^{3}$ &  1.6652407610$\times 10^{3}$\\
TL04 &          $p_{\rm 30}$ &  1.6220939110$\times 10^{3}$ &  1.6652408790$\times 10^{3}$\\
CDM94 &&  1.6220938720$\times 10^{3}$ &  1.6652410890$\times 10^{3}$\\
\hline
This Work &&  7.7357674655$\times 10^{2}$ &  8.0228405811$\times 10^{2}$\\
TL04 &          $p_{\rm 20}$ &  7.7357675130$\times 10^{2}$ &  8.0228406290$\times 10^{2}$\\
CDM94 &&  7.7357672120$\times 10^{2}$ &  8.0228405760$\times 10^{2}$\\
\hline
This Work &&  2.3362818835$\times 10^{2}$ &  2.4860043045$\times 10^{2}$\\
TL04 &          $p_{\rm 10}$ &  2.3362820270$\times 10^{2}$ &  2.4860044040$\times 10^{2}$\\
CDM94 &&  2.3362818910$\times 10^{2}$ &  2.4860043330$\times 10^{2}$\\
\hline
This Work &&  1.5263662310$\times 10^{1}$ &  1.8443609723$\times 10^{1}$\\
TL04 &           $p_{\rm 1}$ &  1.5263662338$\times 10^{1}$ &  1.8443608440$\times 10^{1}$\\
CDM94 &&  1.5263660431$\times 10^{1}$ &  1.8443605420$\times 10^{1}$\\
\hline
This Work &&  8.1753397221 &  9.4137919393\\
TL04 &           $f$ &  8.1753397230 &  9.4137926170\\
\hline
This Work &&  4.9145734152 &  6.7669725650\\
TL04 &          $g_{\rm 1}$ &  4.9145734160 &  6.7669720220\\
CDM94 &&  4.9145731920 &  6.7669711110\\
\hline
This Work &&  3.2249531558$\times 10^{-1}$ &  5.8751831508$\times 10^{-1}$\\
TL04 &         $g_{\rm 10}$&  3.2249534130$\times 10^{-1}$ &  5.8751833439$\times 10^{-1}$\\
CDM94 &&  3.2249531150$\times 10^{-1}$ &  5.8751821518$\times 10^{-1}$\\
\hline
This Work &&  9.7498882680$\times 10^{-2}$ &  1.8574303847$\times 10^{-1}$\\
TL04 &         $g_{\rm 20}$&  9.7498878837$\times 10^{-2}$ &  1.8574304879$\times 10^{-1}$\\
CDM94 &&  9.7498863935$\times 10^{-2}$ &  1.8574303389$\times 10^{-1}$\\
\hline
This Work &&  4.6535316784$\times 10^{-2}$ &  9.0075142564$\times 10^{-2}$\\
TL04 &         $g_{\rm 30}$&  4.6535320580$\times 10^{-2}$ &  9.0075135231$\times 10^{-2}$\\
CDM94 &&  4.6535316855$\times 10^{-2}$ &  9.0075127780$\times 10^{-2}$\\
\hline
This Work &&  2.7186954918$\times 10^{-2}$ &  5.3050607114$\times 10^{-2}$\\
TL04 &         $g_{\rm 40}$&  2.7186956257$\times 10^{-2}$ &  5.3050611168$\times 10^{-2}$\\
CDM94 &&  2.7186954394$\times 10^{-2}$ &  5.3050607443$\times 10^{-2}$
\enddata
\label{Tab:polytrope_n3_EigenfreqAdiabatic}
\end{deluxetable}
In this section, we compare our fiducial eigenfrequencies calculated in the previous section with results in the literature which rely both on the Riccati method (TL04) and on standard relaxation and shooting techniques (CDM94).  
Our fiducial eigenfrequencies are listed in Table~\ref{Tab:polytrope_n3_EigenfreqAdiabatic}, together with the eigenfrequencies calculated by TL04 and CDM94. In Fig.~\ref{fig:goldenComparisonWithPreviousResults} we show the relative difference between our fiducial values and TL04 and CDM94 results (with filled circles), and the relative difference between TL04 and CDM94 (with ``$\times$''). 
The relative difference between our fiducial values and TL04 (CDM94) is between $\sim 10^{-11}-10^{-8}$ ($\sim 10^{-8}-10^{-7}$) both for $l\,=\,2$ and $l\,=\,3$. The upper end of this intervals agrees with the relative difference between TL04 and CDM94. In particular, the better agreement with TL04, results in a relative difference between our fiducial values and CDM94 that is nearly the same as the one between TL04 and CDM94 (orange filled circles overlap with ``$\times$''). This behavior occurs for both the $l\,=\,2$ and $l\,=\,3$ eigenfrequencies.
Among the factors that might be contributing to the small differences between the various results presented here are: the different accuracies adopted during the calculation of the polytropic model among the different studies, the accuracy adopted during the calculation of the eigenfrequencies, and round-off errors, as suggested by TL04 ( \S~5 of their paper).

Hence forward we focus on the harmonic degree $l\,=\,2$. 
\begin{deluxetable}{cc|cc}[!h]
\tablecolumns{2}
\tablewidth{0pc}
\tablecaption{Orthogonality of the $l\,=\,2$ eigenfunctions for a $n_{\rm p}$~=~3 polytropic model. The coefficient $|h_{\rm 0}|$ for the $f-$mode is 1. The eigenfunctions were computed with the integrators accuracies set to $10^{-12}$.}
\tablehead{
\colhead{mode} & \colhead{$|h_{\rm 0}|$}&\colhead{mode} & \colhead{$|h_{\rm 0}|$}}
\startdata
$p_5$  & 1.1~$\times~10^{-6}$&$g_1$  & 5.0~$\times~10^{-6}$\\
$p_4$  & 6.3~$\times~10^{-7}$&$g_2$  & 5.4~$\times~10^{-6}$\\
$p_3$  & 2.5~$\times~10^{-6}$&$g_3$  & 2.3~$\times~10^{-6}$\\
$p_2$  & 9.6~$\times~10^{-6}$&$g_4$  & 1.7~$\times~10^{-6}$\\
$p_1$  & 3.2~$\times~10^{-5}$&$g_5$  & 1.3~$\times~10^{-6}$
\enddata
\label{Tab:polytrope_n3_l2_orthogonality}
\end{deluxetable}
\subsubsection{Testing {\tt CAFein} on a Zero Age Main Sequence Star: the Non-Adiabatic Case}\label{Testing CAFein on a Zero Age Main Sequence Star: the Non Adiabatic Case}

In this section we test the behavior of {\tt CAFein} in the non-adiabatic regime by applying it to a Zero Age Main Sequence (ZAMS) model in the $\beta$ Cephei/SPB region of the HR diagram. Our goal here is only to verify the reliability of {\tt CAFein} in identifying unstable modes; a detailed study of $\beta$ Cepheis and SPBs is beyond the scope of this work. We refer to, e.g., \cite{GautschySaio1995, GautschySaio1996} for a review of pulsating stars and to \cite{CMRI1992}, \cite{MoskalikDziembowski1992}, \cite{KEEG1992}, and \cite{DziembowskiMP1993} for detailed investigations targeting $\beta$ Cephei and SPB variables.

Insofar as the stellar model adopted here is concerned, we follow \cite{SaioCox1980} (hereafter SC80), who investigated $\beta$ Cepheis near the MS and originally found all the models investigated to be stable. Stability was due to the use of opacity formulae prior to the one proposed by \cite{RogersIglesias1992}. Only after the new OPAL opacity tables \citep{RogersIglesias1992, IRW1992, SYMP1994, KEEG1992, MoskalikDziembowski1992} were introduced did it became clear that the excitation mechanism responsible for pulsations in these stars is the so-called $\kappa$-mechanism due to an opacity bump in the heavy elements. This is in contrast to stars located in the classical instability strip whose oscillations are driven by the $\kappa$-mechanism due to partial ionization of H and He I and/or He II.

Following SC80, we use MESA \citep{PBDHLT2011} to create a ZAMS model of 7~$M_\odot$ at metallicity Z~=~0.03 and X~=~0.7. 
We then increase the number of mesh points by interpolating the model with a third order polynomial \citep{Steffen1990}, to reach a resolution between two adjacent mesh points of $\Delta r \leq 5\times 10^{-5}$. Our stellar model has a luminosity, effective temperature, and radius of log($L/{\rm L}_\odot$)~=~3.249, log($T_{\rm eff}/{\rm K}$)~=~4.309, and log($R/{\rm R}_\odot$)~=~0.530, respectively. The H abundance at the center is 0.67. For comparison, the same properties for one of the models used by SC80 are log($L/{\rm L}_\odot$)~=~3.246, log($T_{\rm eff}/{\rm K}$)~=~4.311, and log($R/{\rm R}_\odot$) ~=~0.523, while the H abundance at the center is 0.7.
We conveniently pass onto dimensionless quantities by expressing the physical quantities in the units listed in Table~\ref{Tab:unitsOfPhysicalQuantities}. 

\begin{figure}[t!]
\epsscale{1.1}
\plotone{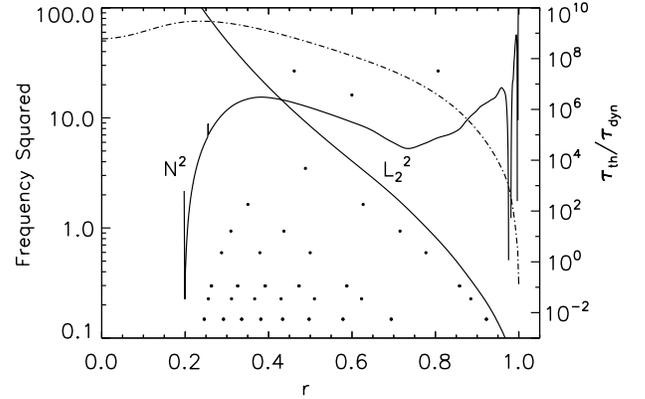}
\caption{Propagation diagram and degree of non-adiabaticity for the 7~$M_\odot$ ZAMS star described in the text. {\textit{Left y-axis}}: the solid lines denote the Brunt-V$\ddot{\rm a}$is$\ddot{\rm a}$l$\ddot{\rm a}$ ($N$) and Lamb ($L_{\rm l}$) frequencies squared, while the dots represent the zeros of the radial part of the eigenfunctions of the modes (from the top to the bottom) $p_{\rm 2}, p_{\rm 1}, g_{\rm 1}$-$g_{\rm 9}$. The harmonic degree was set to $l$~=~2 and the units in Table~\ref{Tab:unitsOfPhysicalQuantities} are used. {\textit{Right y-axis}}: the dashed-dot line shows ratio of the thermal timescale ($\tau_{\rm th}$) to the dynamical timescale ($\tau_{\rm dyn}$). Non-adiabatic effects become significant when the two timescales are comparable (see text).}
\label{fig:7MsunZAMS_propagationDiag_nonAdiabaticity} 
\end{figure}

To have a sense of where the modes can propagate inside the star and whether non-adiabatic effects are significant, we calculate a few adiabatic eigenfrequencies and place them on the propagation diagram (see Fig.~\ref{fig:7MsunZAMS_propagationDiag_nonAdiabaticity}).
This diagram shows that $p-$modes can propagate all the way to the surface, while $g-$modes are more confined towards the star's interior, especially the low-order modes. To investigate the importance of non-adiabatic effects for this stellar model and for the modes of interest, we consider the ratio of the star's thermal to dynamical timescale ($\tau_{\rm th}/\tau_{\rm dyn}$). Since the fundamental oscillation timescale (defined as the travel time of a sound wave from the center to the surface) is of the same order of magnitude as $\tau_{\rm dyn}$, non-adiabatic effects are relevant when $\tau_{\rm th}/\tau_{\rm dyn}$ is small (see e.g. \citealt{UnnoEtAl1989}). The dashed-dot line in Fig.~\ref{fig:7MsunZAMS_propagationDiag_nonAdiabaticity} shows that even though  $\tau_{\rm th}>>\tau_{\rm dyn}$ through the bulk of the star, non-adiabaticity becomes relevant approaching the surface, where $p$-modes can propagate.

\begin{deluxetable}{c|c|c|c}
\tablecolumns{3}
\tablewidth{0pc}
\tablecaption{Non-adiabatic $l\,=\,2$ eigenfrequencies for the 7$M_\odot$ ZAMS described in the text. }
\tablehead{
\colhead{mode} & \colhead{$\omega_{\rm R}$}&\colhead{$\omega_{\rm I}$}}
\startdata
$p_2$& 5.1635 &  1.02~$\times~10^{-4}$ \\
$p_1$& 4.01794 &   -5.86~$\times~10^{-7}$\\
$f$ & 3.18136      &   -2.13~$\times~10^{-7}$\\
$g_1$& 1.86456 &   -1.71~$\times~10^{-7}$\\
$g_2$& 1.27851 &   -3.75~$\times~10^{-7}$\\
$g_3$& 9.66801~$\times~10^{-1}$& -5.85~$\times~10^{-7}$\\
$g_4$& 7.71481~$\times~10^{-1}$& -7.92~$\times~10^{-7}$\\
$g_5$& 6.39310~$\times~10^{-1}$& -9.66~$\times~10^{-7}$\\
$g_6$& 5.45175~$\times~10^{-1}$& -1.06~$\times~10^{-6}$\\
$g_7$& 4.76419~$\times~10^{-1}$& -9.79~$\times~10^{-7}$\\
$g_8$& 4.25101~$\times~10^{-1}$& -4.20~$\times~10^{-7}$\\
$g_9$& 3.84928~$\times~10^{-1}$&  1.11~$\times~10^{-6}$
\enddata
\label{Tab:MESA_ZAMS_7Msun_0.3Zsun_Xc0.67_log47_eigenfreq_nonAd}
\end{deluxetable}

\begin{figure}[h!]
\epsscale{1}
\plotone{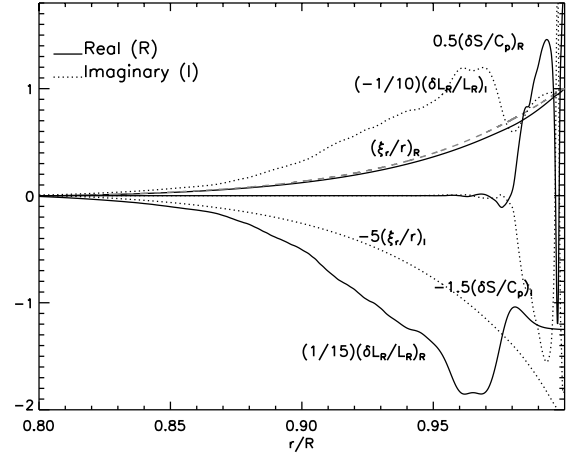}
\caption{Non-adiabatic $l$~=~2 $p_{\rm 2}$-mode for the 7~$M_\odot$ ZAMS star described in the text. Some of the eigenfunctions are shown as a function of the normalized stellar radius close to the surface. The eigenfunctions have been normalized so that $\xi_{\rm r}/r~=~(\xi_{\rm r}/r)_{\rm R} + i(\xi_{\rm r}/r)_{\rm I} = 1$ at the star's surface. Recalling that $y_{\rm 1} = \xi_{\rm r}/r$ and expressions (\ref{eq:IntroducingY1Y2})-(\ref{eq:IntroducingY5Y6}), this normalization yields for each eigenfunction $y$, $(y_{\rm 1,R}y_{\rm R}+y_{\rm 1,I}y_{\rm I})/(y_{\rm 1,R}^{2}+y_{\rm 1,I}^{2}) \rightarrow y_{\rm R} $ and $(y_{\rm 1,R}y_{\rm I}-y_{\rm 1,I}y_{\rm R})/(y_{\rm 1,R}^{2}+y_{\rm 1,I}^{2}) \rightarrow y_{\rm I} $, where the real and imaginary part of $y_{\rm 1}$ are evaluated at the surface. Solid and dotted lines represent the real and imaginary part, respectively. The adiabatic radial component of the displacement from the equilibrium position is also shown for comparison (dashed line).}
\label{fig:7MsunZAMS_p2_nonAdiabatic_SaioLike} 
\end{figure}

Next, we follow the procedure outlined in \S~\ref{The Calculation of the Eigenfrequencies} and calculate the non-adiabatic eigenfrequencies. 
As the inclusion of non-adiabatic effects renders the stellar pulsation equations stiff, at this stage of the calculation we use a variable step implicit Bulirsch-Stoer integrator with an accuracy requirement of $10^{-13}$. 
The results are summarized in Table~\ref{Tab:MESA_ZAMS_7Msun_0.3Zsun_Xc0.67_log47_eigenfreq_nonAd}.
From a test on {\tt CAFein}'s numerical stability like the one performed in \S~\ref{Testing CAFein numerical accuracy on a Polytrope: the Adiabatic Case}, we find that the calculated eigenfrequencies are accurate as long as the required integrator accuracy (both absolute and relative) is $\lesssim 10^{-12}$.
As a negative $\omega_{\rm I}$ denotes an unstable mode, we can see the excitation of the modes which lie in the transition region between $g-$ and $p-$modes, as expected for $\beta$ Cepheis (e.g. \citealt{GautschySaio1996}) and SPBs. The latter are generally understood as an extension of the $\beta$ Cephei instability towards longer periods (smaller frequencies), as their observed pulsation periods are due to the excitation of $g-$modes.
As a reference, the periods for the $p_{\rm 1}-$  and $g_{\rm 8}-$modes listed in Table~\ref{Tab:MESA_ZAMS_7Msun_0.3Zsun_Xc0.67_log47_eigenfreq_nonAd} in days are $\simeq$~ 0.07 and $\simeq$~ 0.6, respectively, which is consistent with the range of oscillation periods observed for these kind of stars.
Here we also note that the magnitude of $\omega_{\rm I}$ (and therefore non-adiabaticity) is negligible for $g-$modes, while it increases by about two order of magnitudes for the $p_{\rm 2}-$mode. This was expected given the trend of  $\tau_{\rm th}/\tau_{\rm dyn}$ shown in Fig.~\ref{fig:7MsunZAMS_propagationDiag_nonAdiabaticity}. 
Since the $p_{\rm 2}-$mode is the only mode considered here for which dissipation is significant, we calculate the non-adiabatic eigenfunctions for this mode. Recall that the calculation of the eigenfunctions is performed in two steps. A first integration yields the components of the Riccati matrix and the permutations applied, while a second integration uses these information for the calculation of the eigenfunctions. During the first integration {\tt CAFein} uses a Runge-Kutta integrator with an accuracy requirement of $10^{-14}$. During the calculation of the eigenfunctions we integrate Eq.~(\ref{eq:dvEigenfunc}) using again a Runge-Kutta integrator with an accuracy requirement of $10^{-13}$.
Following the representation of the $f-$mode eigenfunctions in Fig. ~1 of SC80, we show in Fig.~\ref{fig:7MsunZAMS_p2_nonAdiabatic_SaioLike} some of the non-adiabatic eigenfunctions for the $p_{\rm 2}$-mode.
The purely adiabatic radial component of the displacement from the equilibrium position (dashed line) is also shown for comparison. As expected, $(\xi_{\rm r}/r)_{\rm R}$ and its adiabatic counterpart are very similar and the entropy perturbation [both $(\delta S/c_{\rm p})_{\rm R}$ and $(\delta S/c_{\rm p})_{\rm I}$] increases rapidly towards the star's surface, where non-adiabaticity becomes significant. Similarly to the results presented by SC80, the real part of the entropy perturbation presents two minima which are located at the peaks of the opacity $\kappa$. For the case of SC80, the peaks are in the He ionization zone. In our MESA model, the lower-temperature bump in the opacity is due to the He ionization, while the one at a higher temperature is due to photon absorption by the $L$-shell of Fe and photoionization from the $K$-shell of C, O, and Ne \citep{RogersIglesias1992}. It is the $k-$mechanism associated with this second bump at a temperature of $\simeq 2~\times~10^{5}$~K which drives the pulsations observed in $\beta$ Cephei variables.

The purpose of this section was to prove that {\tt CAFein} can identify unstable modes in a $\beta$ Cephei/SPB variable star, if the OPAL opacity tables \citep{RogersIglesias1992, IRW1992, SYMP1994, KEEG1992, MoskalikDziembowski1992} are used. However, we neglected the effect of convection, which might affect or even excite non-radial pulsations, as mentioned in \S~\ref{The Equations Governing Non-Adiabatic and Non-Radial Stellar Pulsations}. The thin convective regions at the star's surface are visible in the propagation diagram shown in Fig.~\ref{fig:7MsunZAMS_propagationDiag_nonAdiabaticity}, where $N^{2}~\textless~0$.  
\section{Investigating Dynamic Tides With {\tt CAFein}}\label{Investigating Dynamic Tides With CAFein}

Before describing the extension of the Riccati method developed to investigate tidally excited stellar oscillations, we briefly outline the basic assumptions adopted in this work and introduce the various parameters entering the dynamic tides theoretical framework. 

\subsection{The Tide-Generating Potential}\label{The Tide-Generating Potential}
We consider a close binary system of stars with masses $M_{\rm 1}$ (primary) and $M_{\rm 2}$ (secondary) orbiting around one another in a Keplerian orbit. 
We assume that the primary has a radius $R_{\rm 1}$ and that it rotates uniformly around an axis orthogonal to the orbital plane with angular velocity $\Omega_{\rm 1}$ in the sense of the orbital motion, while we treat the companion as a point mass. Furthermore, we assume $\Omega_{\rm 1}$ to be small enough so that the Coriolis force and the centrifugal force can be neglected. Under these assumptions, the tides raised by the companion can be treated as small forced perturbations applied on a spherically symmetric star in hydrostatic equilibrium. 
Following the general procedure, we can express the tide-generating potential in spherical coordinates ${\bf r} = (r, \theta, \phi)$ with respect to an orthogonal frame corotating with the star and expand it in Fourier series as (e.g. \citealt{PolflietSmeyers1990}, hereafter PS90)
\begin{small}
\begin{align}
\epsilon_{\rm T}W({\bf r}, t) = & -\epsilon_{\rm T}\sum_{l=2}^{4}\sum_{m=-l}^{l}\sum_{k=-\infty}^{\infty}c_{l,m,k}\left(\frac{r}{R_{1}}\right)^{l}Y^m_l(\theta, \phi)\nonumber\\
&\times{\rm exp}[i(\sigma_{m,k}t-k\Omega_{\rm orb}\tau)]
\label{eq:tidalPotential}
\end{align}
\end{small}
where the polar angle $\theta$ is measured from the rotational angular velocity vector, while the azimuthal angle $\phi$ is measured in the orbital plane and in the sense of the orbital motion. At time $t$~=~0, the angle $\phi$~=~0 marks the position of the periastron of the binary orbit. The tide-generating potential is a solution to Laplace's equation. The indices $l, m$, and $k$ in Eq.~(\ref{eq:tidalPotential}) are the harmonic degree, the azimuthal number, and the Fourier index, respectively. The dimensionless parameter $\epsilon_{\rm T} \equiv(R_{\rm 1}/a)^3(M_{\rm 2}/M_{\rm 1})$ measures the ratio of the tidal force to gravity at the star's equator, $a$ is the semi-major axis, $\sigma_{m,k} = k\Omega_{\rm orb}+m\Omega_{\rm 1}$ is a forcing angular frequency with respect to the corotating frame, $\Omega_{\rm orb}~=~2\pi/P_{\rm orb}$ the mean motion, $\tau$ a time at periastron passage, and $c_{l,m,k}$ are Fourier coefficients defined as
\begin{small}
\begin{align}
c_{l,m,k}  = &\frac{(l-|m|)!}{(l+|m|)!}P^{|m|}_{l}(0)\left(\frac{R_{\rm 1}}{a}\right)^{l-2}\frac{1}{(1-e^{\rm 2})^{l-1/2}}\frac{1}{\pi}\nonumber\\
&\times\int_0^\pi(1+e~{\rm cos}\nu)^{\rm l-1}{\rm cos}(kM+m\nu)d\nu.
\label{eq:clmk}
\end{align}
\end{small}
Here, $P^{m}_{l}(cos\theta)$ are Legendre polynomials of the first kind, $\nu$ is the true anomaly and $M~=~\Omega_{\rm orb}(t-\tau)$ the mean anomaly. The main properties of the Fourier coefficients were described by \cite{SWVH1998}, PS90, and \cite{WDK2010}. Briefly, $c_{l,m,k}$ are symmetric with respect to $m$ and $k$ ($c_{l,m,k}~=~c_{l,-m,-k}$) and are equal to zero for odd values of  $l+|m|$ since $P^{m}_{l}(0)=0$ for odd values of  $l+|m|$. Furthermore, the binomial theorem implies that $c_{l,m,0}~=~0$. 
For a given orbital eccentricity, the absolute value of $c_{l,m,k}$ decreases with increasing $k$, though the decrease is slower for highly eccentric orbits \citep{Willems2003, WVHS2003, SWVH1998}.
This implies that the number of $c_{l,m,k}$ terms with non-trivial contributions to the tide-generating potential is finite, though it increases with increasing eccentricity.
Given the dependence of $c_{l,m,k}$ on $(R_{\rm 1}/a)^{l-2}$, investigations on dynamic tides are often restricted to the terms belonging to $l$~=~2, as they are dominant.

It is clear from the expansion~(\ref{eq:tidalPotential}) of the tide-generating potential that the tidal action from the companion induces in the primary an infinite number of forcing angular frequencies $\sigma_{m,k}$. The terms associated with $\sigma_{m,k}~=~0$ (the time-independent terms in $\epsilon_{\rm T}W$) give rise to {\textit {static}} tides, while the terms associated with $\sigma_{m,k}~\neq~0$ (the time-dependent terms in $\epsilon_{\rm T}W$) give rise to {\textit {dynamic}} tides. 

In the limit of an infinite orbital period, tides are referred to as {\emph{equilibrium}} tides (e.g. PS90, \citealt{WDK2010})
\subsection{The Equations For Tidally Excited Stellar Pulsations and The Secular Evolution of the Orbital Elements}\label{The Equations For Tidally Excited Stellar Pulsations}
If the tides raised by the companion are treated as a small perturbation applied on a spherically symmetric star in hydrostatic equilibrium, the equations describing forced stellar oscillations are still derived from Eqs.~(\ref{eq;massConserv})-(\ref{eq:radDiffusion}), provided that the $\epsilon_{\rm T}W$ is added to Eq.~(\ref{eq;momConserv}) of momentum conservation (e.g. PS90, \citealt{WDK2010}).
Following the standard procedure, we take the unperturbed solution to be axisymmetric and assume that the spatial and temporal part of a small perturbation can be written in 
Eulerian form as $f'_{\rm T}(r, \theta, \phi, t) = \sum_{l=2}^{4}\sum_{m=-l}^{l}\sum_{k=-\infty}^{\infty}f_{l,m,k}'(r)Y^m_l(\theta, \phi)e^{i\sigma_{m,k} t}$ or similarly for the Lagrangian form, denoted with $\delta$.
Since the tide-generating potential is a solution to Laplace's equations, perturbing and linearizing the basic Eqs.~(\ref{eq;massConserv})-(\ref{eq:radDiffusion}) with the new equation for momentum conservation yields, for each set of ($l,m,k$) in the expansion of the tide-generating potential, a system of equations which is formally identical to Eqs.~(\ref{eq:y1})-(\ref{eq:y6}), with the following modifications.
The perturbation of the star's gravitational potential and the tide-generating potential are grouped into the total perturbation of the gravitational potential defined as $\Psi = \Phi_{\rm T}' + \epsilon_{\rm T}W$ (e.g. \citealt{Zahn1975}, PS90, \citealt{WDK2010}), 
where $\Phi_{\rm T}'$ (denoted in \S~\ref{The Equations Governing Non-Adiabatic and Non-Radial Stellar Pulsations} as $\Phi'$) is the perturbation of the star's potential of self-gravitation due to  the tidal action of the companion. Furthermore, the new integration variables retain the same form, provided that $\Phi'$ is substituted with $\Psi$.
A final modification to Eqs.~(\ref{eq:y1})-(\ref{eq:y6}) concerns the BCs at the star's surface. As the gravitational potential and its first derivative must be continuous at $r~=~R_{\rm 1}$, BC~(\ref{eq:BCsurf2}) becomes (e.g. PS90)
\begin{align}
& y_{\rm 4}+(l+1)y_{\rm 3}+\frac{4\pi \rho}{g} y_{\rm 1}+\frac{\epsilon_{T}(2l+1)c_{lmk}}{g}=0
\label{eq:BCsurf2tides}
\end{align}
where $\rho$ and $g$ are made dimensionless via the units listed in Table~\ref{Tab:unitsOfPhysicalQuantities}.
Therefore, the introduction of $\Psi$ keeps the tidally excited stellar pulsation Eqs.~(\ref{eq:y1})-(\ref{eq:y6}) homogeneous, but it renders the BCs non-homogeneous. Because of the non-homogeneous term in Eq.~(\ref{eq:BCsurf2tides}), the solutions to Eqs.~(\ref{eq:y1})-(\ref{eq:y6}) are proportional to $\epsilon_{\rm T}c_{l,m,k}$. Furthermore, even though the system of equations is complex, the dimensionless tidal forcing frequency $\omega_{m,k}$,  is purely real (recall that $\omega^{2}_{m,k}~=~\sigma^{2}_{m,k} R^{3}(GM)^{-1}$). In what follows, we refer to the solution of Eqs.~(\ref{eq:y1})-(\ref{eq:y6}) with BCs~(\ref{eq:BCcenter1})-(\ref{eq:BCcenter3}) at the star's center, and BCs~(\ref{eq:BCsurf1}), (\ref{eq:BCsurf2tides}), and (\ref{eq:BCsurf3}) at the star's surface with ``tidal eigenfunctions''.

From the tidal eigenfunctions, the timescales for the secular evolution of the orbital elements and stellar spin due to dynamic tides can be readily calculated. 

The evolution of the orbital separation and eccentricity is due to the primary's tidal deformation, which in turn perturbs the external gravitational field and therefore the Keplerian motion of the binary components. 
Energy dissipation in the surface layers causes a phase shift between the perturbation of the gravitational potential and the companion's position. This phase shift results in a torque exerted from the secondary on the tidally deformed primary, which affects the primary's spin.
The rates of secular evolution for $a$, $e$, and $\Omega_{\rm 1}$ are given by (e.g. \citealt{WDK2010})
\begin{small}
\begin{align}
\left(\frac{da}{dt}\right)_{\rm sec} = &\frac{8\pi}{P_{\rm orb}}\frac{M_{\rm 2}}{M_{\rm 1}}a \sum_{l=1}^{4}\sum_{m=-l}^{l}\sum_{k=0}^{\infty}\left(\frac{R_{\rm 1}}{a}\right)^{l+3}\nonumber\\
&\times\kappa_{l,m,k}|F_{l,m,k}|{\rm sin}\gamma_{l,m,k}G^{(2)}_{l,m,k}(e)\label{eq:dadt_tides_final}\\
\left(\frac{de}{dt}\right)_{\rm sec} = & \frac{8\pi}{P_{\rm orb}}\frac{M_{\rm 2}}{M_{\rm 1}}\sum_{l=1}^{4}\sum_{m=-l}^{l}\sum_{k=0}^{\infty}\left(\frac{R_{\rm 1}}{a}\right)^{l+3}\nonumber\\
&\kappa_{l,m,k}|F_{l,m,k}|{\rm sin}\gamma_{l,m,k}G^{(3)}_{l,m,k}(e)\label{eq:dedt_tides_final}\\
\left(\frac{d\Omega_{\rm 1}}{dt}\right)_{\rm sec} = & \frac{8\pi}{P_{\rm orb}}\left(\frac{GM_{\rm 1}^{2}M_{\rm 2}^{2}}{M_{\rm 1}+M_{\rm 2}}\right)^{1/2}\frac{M_{\rm 2}}{M_{\rm 1}}\frac{a^{1/2}}{I_{\rm 1}} \nonumber\\
&\sum_{l=1}^{4}\sum_{m=-l}^{l}\sum_{k=0}^{\infty}\left(\frac{R_{\rm 1}}{a}\right)^{l+3}\kappa_{l,m,k}|F_{l,m,k}|\nonumber\\
&\times{\rm sin}~\gamma_{l,m,k}G^{(4)}_{l,m,k}(e)
\label{eq:dOmegadt_tides_final}
\end{align}
\end{small}
where $I_{\rm 1}$ is the star's moment of inertia and Eq.~(\ref{eq:dOmegadt_tides_final}) is derived assuming solid-body rotation and that the tidal deformation does not affect $I_{\rm 1}$. In the above equations, the dimensionless $F_{l,m,k}$ measure the response of the star to the various tidal forcing frequencies and are given by
\begin{equation}
F_{l,m,k} = -\frac{1}{2}\left[\frac{R_{1}}{GM_{1}}\frac{\Psi_{l,m,k}(R_{\rm 1})}{\epsilon_{\rm T}c_{l,m,k}}+1\right].
\label{eq:Flmk}
 \end{equation}
These coefficients are independent of $\epsilon_{\rm T}c_{l,m,k}$ because $\Psi_{l,m,k}(R_{\rm 1})\propto\epsilon_{\rm T}c_{l,m,k}$ (\S~\ref{The Equations For Tidally Excited Stellar Pulsations}). 
In the units of Table~\ref{Tab:unitsOfPhysicalQuantities}, expression~(\ref{eq:Flmk}) reduces to
\begin{equation}
F_{l,m,k} = -\frac{1}{2}\left[\frac{(y_{\rm 3})_{l,m,k}(1)g(1)}{\epsilon_{\rm T}c_{l,m,k}}+1\right] \equiv |F_{l,m,k}|e^{i\gamma_{l,m,k}}
\label{eq:FlmkCAFeinUnits}
 \end{equation}
where the last equality comes from the complex nature of the tidal eigenfunctions. For the various properties of symmetry obeyed by $|F_{l,m,k}|$ and for the definition of $\kappa_{l,m,k}$ (not to be confused with the opacity), $G^{(2)}_{l,m,k}(e), G^{(3)}_{l,m,k}(e)$, and $G^{(4)}_{l,m,k}(e)$, and their properties we refer to \cite{WVHS2003, WDK2010}. Here we just note that $G^{(3)}_{l,m,k}(e)$ are all zero for a binary with a circular orbit. Eqs.~(\ref{eq:dadt_tides_final})~-~(\ref{eq:dOmegadt_tides_final}) take the same form as the equations for the rate of secular change of orbital separation, eccentricity, and spin derived by \cite{Zahn1977,Zahn1978}, \cite{Hut1981}, and \cite{Ruymaekers1992}, in the limiting case of weak damping and small forcing angular frequencies (see Appendix D of \citealt{WDK2010} for a derivation).

In what follows, we omit the subscripts $l$, $m$, and $k$ from the components of the tidal displacement field and the perturbed stellar structure quantities and we denote the tidal forcing frequency by $\omega_{\rm T}$.
\subsection{Extending the Riccati Method to Forced Stellar Oscillations}\label{Extending the Riccati Method to Forced Stellar Oscillations}

As described in \S~\ref{The Riccati Method}, the Riccati method relies in going from a homogeneous system of ordinary differential equations to a \emph{non-homogeneous} one. 
However, as explained in \S~\ref{The Equations For Tidally Excited Stellar Pulsations}, even though the equations describing the tidally excited stellar oscillations are homogeneous, the same is not true for the BCs at the star's surface.
We make the Riccati method viable for investigating dynamic tides by introducing two new variables $y_{\rm 7}$ and $y_{\rm 8}$, such that BC~(\ref{eq:BCsurf2tides}) at the star's surface becomes homogeneous
\begin{equation}
y_{\rm 4}+(l+1)y_{\rm 3}+\frac{4\pi \rho}{g} y_{\rm 1}+\frac{\epsilon_{T}(2l+1)c_{lmk}}{g}y_{\rm 8}=0
\label{eq:BCsurf2tidesy8}
\end{equation}
The introduction of two variables instead of one is required to keep all the matrices entering the Riccati method square. We take $y_{\rm 7}$ and $y_{\rm 8}$ to be a solution of the following differential equation
\begin{equation}
r\frac{dy_{\rm 7}}{dr} = r\frac{dy_{\rm 8}}{dr} = 0
\label{eq:y7y8}
\end{equation}
with BCs at the center and at the surface given by
\begin{equation}
y_{\rm 7}~=~y_{\rm 8}.
\label{eq:BCy7y8}
\end{equation}
Once the tidal eigenfunctions are determined, we normalize them so that $y_{\rm 8}~=~1$ at the star's surface. This choice of normalization causes Eq.~(\ref{eq:BCsurf2tidesy8}) to reduce to the original BC~(\ref{eq:BCsurf2tides}).
Here we note that changing the form of Eq.~(\ref{eq:y7y8}) does not affect our results, but it can affect the running time.
With the introduction of $y_{\rm 7}$ and $y_{\rm 8}$, the new definition of vectors ${\bf u_{\rm Ric}}$ and ${\bf v_{\rm Ric}}$ (see \S~\ref{The Riccati Method}) at the star's boundaries becomes
\begin{equation}
{\bf u_{\rm Ric}} = \left(
\begin{array}{c}
 {y_{\rm 1}} \\
 {y_{\rm 4}} \\
 {y_{\rm 5}} \\
 {y_{\rm 7}} 
\end{array}
\right)~~,~~ 
{\bf v_{\rm Ric}} = \left(
\begin{array}{c}
 {y_{\rm 2}} \\
 {y_{\rm 3}} \\
 {y_{\rm 6}} \\
 {y_{\rm 8}} 
\end{array}
\right)\label{eq:initialUVTides}
\end{equation}
and the new initial conditions on the Riccati matrices at the star's center and surface are derived accordingly. 
Even though the introduction of the new variables increases the size of the various matrices, it does not affect the running time significantly.
In particular, as far as reimbedding (\S~\ref{Reimbedding}) is concerned, during the search for the permutation yielding ${\bf R'}$ with the minimum Eucledian norm, $y_{\rm 7}$ and $y_{\rm 8}$ are kept fixed. This trick yields the same number of trials as in the non-adiabatic stellar pulsation problem.

The tidal eigenfunctions are calculated as described in \S~\ref{The Calculation of the Eigenfunctions}.
\subsection{Testing the Extension of the Riccati Method to Investigate Dynamic Tides}\label{Testing the Extension of the Riccati Method to Investigate Dynamic Tides}
Here we test whether our extension of the Riccati method to treat tidally excited stellar pulsation works as expected. First, we compare the tidal eigenfunctions calculated with {\tt CAFein} with the work presented by PS90 in the adiabatic regime. Next, we compare the orbital and spin evolution timescales due to dynamic tides (\S~\ref{The Equations For Tidally Excited Stellar Pulsations}) computed with {\tt CAFein} with the results presented by \cite{WVHS2003} (WVHS03, hereafter). Finally, we compute these same timescales for a binary hosting a 1.5$\, M_{\odot}$ Main Sequence star and a hot Jupiter. The purpose of this last exercise is to demonstrate that {\tt CAFein} can handle the high-order modes involved in these binaries.
\subsubsection{Tidal Eigenfunctions of a 5$\,M_{\odot}$ MS star}\label{Tidal Modulation of an Eigenfunction of a 5Msun MS star}
Here, we investigate the effect of non-adiabatic dynamic tides on a MS star of 5$M_\odot$ (primary) and metallicity Z~=~0.018, by studying the variation of the radial component of the tidal displacement at the surface [$\xi_{r}(1)$] as a function of the tidal forcing frequency. 
Our main goal here is to verify numerically that a dynamical tide can be approximated as the sum of the equilibrium tide and another part reflecting the oscillatory properties of the star itself, in agreement with what asymptotic theories have shown  (e.g. \citealt{Zahn1975,Smeyers1997,SmeyersWillems1998}). The equilibrium tide associated with the limiting case of an infinite orbital period is given by $\xi_{eq} = -\Psi/g$ (e.g. PS90).

As before, we create the stellar model with MESA and increase the number of mesh points as described in \S~\ref{Testing CAFein on a Zero Age Main Sequence Star: the Non Adiabatic Case}.
The radius of the model adopted here is $R_{\rm 1}\,\simeq\,2.62\,R_{\odot}$, and the convective core extends out to $r/R_{\rm 1}\,\simeq\,$0.185. For comparison, these same parameters for the model used by PS90 are $R_{\rm 1}\,\simeq\,2.52\,R_{\odot}$ and $r/R_{\rm 1}\,\simeq\,$0.18, respectively. 
We set ($l, m, k$) = (2,0,1), we take the companion to be a point mass of 5.0$\,M_\odot$, and we set $e$~=~0.4.
We then vary the orbital period so that the tidal forcing frequency becomes comparable to the $g_{\rm 9}$- and $g_{\rm 10}$-mode frequencies. Here we note that, since $m\,=\,0$ and $\sigma_{m,k} = k\Omega_{\rm orb}+m\Omega_{\rm 1}$, the star's spin does not need to be specified.

\begin{figure} [!h]
  \epsscale{1}
\plotone{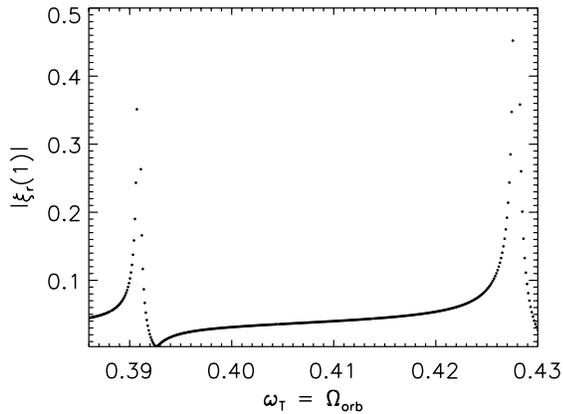}
\caption{Modulus of the radial component of the tidal displacement at the surface for the MS star of 5$M_\odot$ described in the text. The peaks correspond to a resonances with the modes $g_{\rm 9}$ and $g_{\rm 10}$. For the case under consideration, $\omega_{\rm T} = k\Omega_{\rm orb}+m\Omega_{\rm 1} = \Omega_{\rm orb}$ and the frequencies are in the units of Table~\ref{Tab:unitsOfPhysicalQuantities}}
   \label{fig:modXi_r_5Msun_z0p018_polflietSmeyers}
    \end{figure}

The behavior of the radial component of the displacement from the equilibrium position at the surface $|\xi_{\rm r}(1)|$ is shown in Fig.~\ref{fig:modXi_r_5Msun_z0p018_polflietSmeyers} as a function of the tidal forcing frequency (and orbital frequency). This Figure can be compared with Fig.~1 in PS90 or \cite{SavonijePapaloizou1983}. The difference between the Figures presented by these two studies is that \cite{SavonijePapaloizou1983} performed a non-adiabatic calculation, while in the adiabatic calculation of PS90 $\xi_{r}(1)$ changes discontinuously between $\pm~\infty$. Fig.~\ref{fig:modXi_r_5Msun_z0p018_polflietSmeyers} shows that, close to a resonance, the amplitude of the displacement at the surface varies greatly for small changes in $\omega_{\rm T}$, while away from a resonance changes in the amplitude occur at a much slower rate.

\begin{figure} [!h]
  \epsscale{1}
\plotone{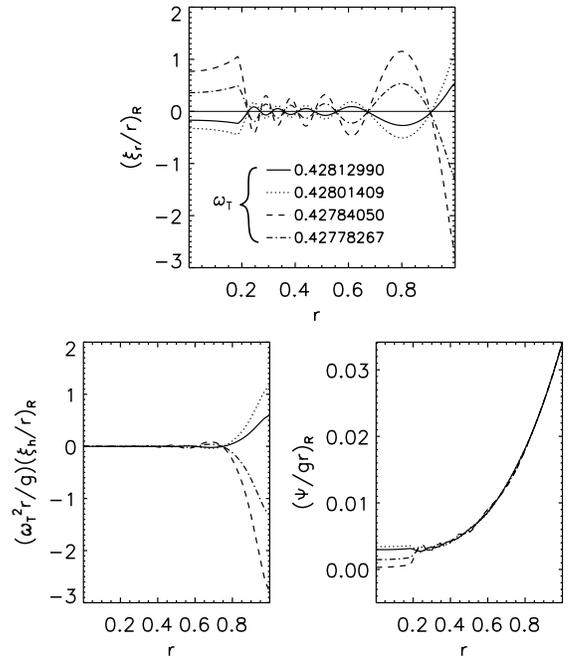}
\caption{Real components of the tidal response for the 5$M_\odot$ MS star described in the text (\S~\ref{Tidal Modulation of an Eigenfunction of a 5Msun MS star}). $\omega_{\rm T}$ is set close to the $g_{\rm 9}$-mode eigenfrequency.}
   \label{fig:y1y2y3_y81atSurf_real_ZAMS_5Msun_log26_g9}
    \end{figure}

In Fig.~\ref{fig:y1y2y3_y81atSurf_real_ZAMS_5Msun_log26_g9} we show the radial and orthogonal components of the tidal displacement and the total perturbation of the gravitational potential as a function of the radial coordinate, for tidal forcing frequencies $\simeq\,0.42778-0.42813$. The nine zeros displayed by $\xi_{\rm r}/r$ indicate that we are close to a resonance with the $g_{\rm 9}$-mode. Our Fig.~\ref{fig:y1y2y3_y81atSurf_real_ZAMS_5Msun_log26_g9} can be compared with Figs.~2 and 3 in PS90, where the discontinuity in $\xi_{r}(1)$ is again a result of the adiabatic treatment.
As PS90 pointed out, the behavior of $(\xi_{\rm r}/r)_{\rm R}$ at the surface is determined by BC~(\ref{eq:BCsurf1}) and, therefore, by the competition of the orthogonal component of the tidal displacement and the total perturbation of the potential. 
This can be seen by considering that, close to a resonance with the $g_{\rm 9}-$mode, $\omega_{\rm T}^{2}\simeq\,0.2$ and $\omega_{\rm T}^{-2}\,\simeq\,$5.5, while the other term $V$ entering BC~(\ref{eq:BCsurf1}) is $V\,\simeq\,1.5\times 10^{\rm 3}$, so $V^{\rm -1}\,\simeq\,6.7\times 10^{\rm -4}$. 
Therefore, for $l$~=~2, it can be shown that BC~(\ref{eq:BCsurf1}) reduces to
\begin{equation}
y_{\rm 1}\,\simeq\,y_{\rm 2}-y_{\rm 3}
\end{equation}

\begin{figure} [h!]
  \epsscale{1.2}
\plotone{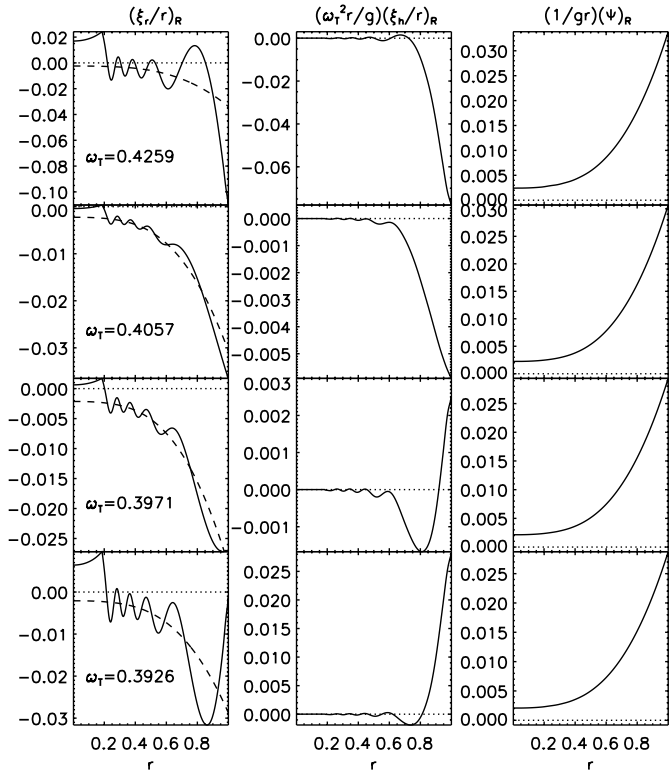}
\caption{Real components of the tidal response for the 5$M_\odot$ MS star described in the text (\S~\ref{Tidal Modulation of an Eigenfunction of a 5Msun MS star}). The horizontal dotted line marks the position of the zero, while the dashed line in the left panels indicates the equilibrium tide component $-\Psi/g$ divided by $r$ (e.g. PS90). In the top and bottom plots, $\omega_{\rm T}$ is close to the $g_{\rm 9}$ and $g_{\rm 10}$ eigenfrequency, respectively (see text).}
   \label{fig:non_adiabatic_eigenfunctions_y81atSurf_PolflietSmeyers90}
    \end{figure}

As shown in Fig.~\ref{fig:y1y2y3_y81atSurf_real_ZAMS_5Msun_log26_g9}, close to resonance the behavior of $\xi_{\rm r}$(1) is related to the trend of $\xi_{\rm h}$(1), as the contribution from the total perturbation of the potential remains small.
Moving along the steep slope in Fig.~\ref{fig:modXi_r_5Msun_z0p018_polflietSmeyers} towards smaller $\omega_{\rm T}$ , the rapid decrease in $\xi_{\rm r}$(1) is related to the rapid decrease of $\xi_{\rm h}$(1) (as PS90 pointed out referring to Fig.~1 in their paper). This can be seen from the decrease in $\xi_{\rm r}$(1) in going from $\omega_{\rm T}\,=\,$0.42778 to 0.4259 (from the dot-dashed line in Fig.~\ref{fig:y1y2y3_y81atSurf_real_ZAMS_5Msun_log26_g9} to the top plot in Fig.~\ref{fig:non_adiabatic_eigenfunctions_y81atSurf_PolflietSmeyers90}, to be compared with the behavior presented by PS90 in going from their Fig.~3 to 4). At $\omega_{\rm T}\,=\,$0.4259 the nine zeros in $\xi_{\rm r}$(r)/r are still visible.
Along the nearly horizontal line of Fig.~\ref{fig:modXi_r_5Msun_z0p018_polflietSmeyers}, as $\omega_{\rm T}$ decreases further, the increasingly significant total perturbation of the potential affects the behavior of $\xi_{\rm r}$(1). As shown in the middle-top panels in Fig.~\ref{fig:non_adiabatic_eigenfunctions_y81atSurf_PolflietSmeyers90} (to compare with Fig.~5 in PS90), at $\omega_{\rm T}$~=~0.4057, $\xi_{\rm r}$(r)/r close to the star's surface shifts from the horizontal dotted line marking the position of $\xi_{\rm r}$(r)/r~=~0. 
Decreasing further the tidal forcing frequency to $\omega_{\rm T}$~=~0.3971 (middle-bottom panel in Fig.~\ref{fig:non_adiabatic_eigenfunctions_y81atSurf_PolflietSmeyers90}), the radial component of the displacement makes a turn upward at the surface (to compare with Fig.~6 in PS90). This behavior is related to the change in sign of $\xi_{\rm h}$(1).
Finally, at $\omega_{\rm T}$~=~0.3926 (bottom panel in Fig.~\ref{fig:non_adiabatic_eigenfunctions_y81atSurf_PolflietSmeyers90}), the contribution from the orthogonal component of the displacement and the total perturbation of the potential become comparable, and the resulting amplitude of $\xi_{\rm r}$(1) is zero (to compare with Fig.~7 in PS90). The radial component of the displacement from equilibrium is approaching the $g_{\rm 10}$-mode. 

From the behavior of $\xi_{\rm r}$(r)/r displayed in Fig.~\ref{fig:non_adiabatic_eigenfunctions_y81atSurf_PolflietSmeyers90}, it is clear that away from resonances a dynamical tide can be approximated as the sum of the equilibrium tide ($-\Psi/g$, denoted with a dashed line) and another part reflecting the oscillatory properties of the star itself, as asymptotic theories have shown (e.g. \citealt{Zahn1975,Smeyers1997,SmeyersWillems1998}).
\subsubsection{Dynamic Tides Timescales in an Eccentric Binary Hosting a 5$\,M_{\odot}$ MS star and a Neutron Star}\label{Dynamic Tides Timescales in an Eccentric Binary Hosting a 5Msun MS star and a Neutron Star}

Here, we test {\tt CAFein}'s results on the orbital and spin evolution timescales due to dynamic tides both in and out of resonance by reproducing the timescales presented by WVHS03 for an eccentric binary hosting a 5$\,M_{\odot}$ MS star (primary) and a neutron star.
Note that, for the calculation of the dynamic tide timescales, WVHS03 did not solve the fully non-adiabatic problem, as we do here, and the perturbed stellar quantities are found via semi-analytical solutions (they do use a full non-adiabatic calculation for the eigenfrequencies used in the semi-analytical solutions).

As before, we use MESA to create a stellar model of a 5$\,M_{\odot}$ MS star at solar metallicity and increase the number of mesh points as described in \S~\ref{Testing CAFein on a Zero Age Main Sequence Star: the Non Adiabatic Case}.
The radius of the model adopted here is $R_{\rm 1}\,\simeq\,2.66\,R_{\odot}$, its dynamical ($\tau_{\rm dyn}$), thermal ($\tau_{\rm th}$) and nuclear ($\tau_{\rm nucl}$) timescales are 323.9\,min, 2.71$\times 10^{5}\,$yr and 9.20$\times 10^{7}\,$yr, respectively. For comparison, these same parameters for the model used by WVHS03 are $R_{\rm 1}\,\simeq\,2.8\,R_{\odot}$, $\tau_{\rm dyn}\,=\,$54.8\,min, $\tau_{\rm th}\,=\,$4.88$\times 10^{5}\,$yr, and $\tau_{\rm nucl}\,=\,$8.67$\times 10^{7}\,$yr, respectively. In both cases the H fraction at the center is $X_{\rm c}\,=\,$0.7.
We take the companion to be a point mass of 1.4$\,M_\odot$, the eccentricity $e$~=~0.5, and we take the spin of the primary to be 50\% of the companion's orbital angular velocity at periastron. 

\begin{figure} [!h]
  \epsscale{1.1}
\plotone{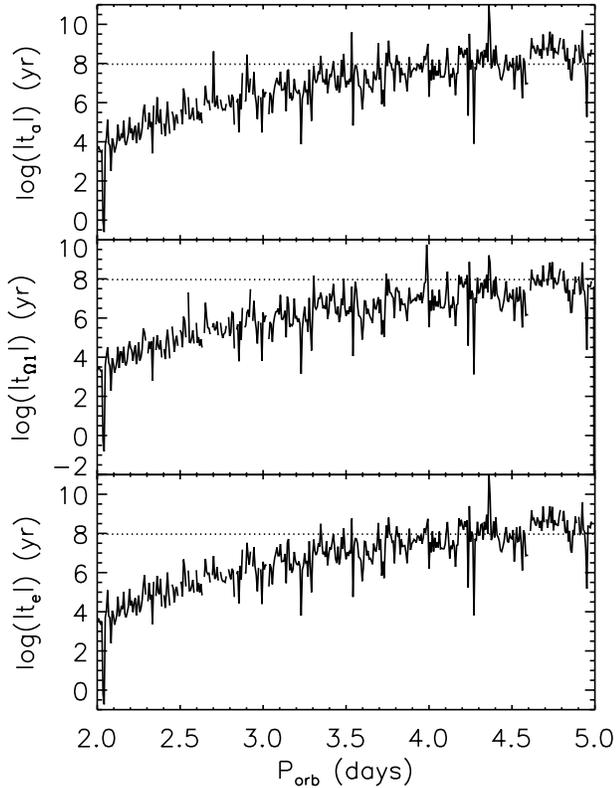}
\caption{Orbital and spin evolution timescales due to dynamic tides for a binary hosting a 5$\,M_{\odot}$ MS star and a neutron star in an eccentric orbit ($e\,=\,0.5$) as a function of the orbital period ($P_{\rm orb}$).{\it Top: } ${\rm log}|t_{a}| = {\rm log}|a/\dot{a}_{\rm sec}|$. {\it Middle: } ${\rm log}|t_{\Omega_{\rm 1}}| ={\rm log}|\Omega_{\rm 1}/\dot{\Omega}_{\rm 1,sec}|$. {\it Bottom: } ${\rm log}|t_{e}| = {\rm log}|e/\dot{e}_{\rm sec}|$. The horizontal dotted line represents the logarithm of the star's nuclear time scale. A comparison with Fig.~3 in WVHS03 indicates satisfactory agreement on the timescale calculation.}
\label{fig:Comparison_WillemEtAl2003_Valsecchi_kUpTo9}
\end{figure}

Similarly to WVHS03, we consider the dominant terms in the expansion of the tide-generating potential and fix $l\,=\,2$ and $m\,=\,-2$. We then calculate the orbital and spin evolution timescales for orbital periods ($P_{\rm orb}$) ranging from 2 to 5\,days, taking into account several terms in the expansion of the tide-generating potential and considering $k$ up to 20 in Eq.~\ref{eq:tidalPotential}. The results are summarized in Fig.~\ref{fig:Comparison_WillemEtAl2003_Valsecchi_kUpTo9}, which should be compared with Fig.~3 in WVHS03 (we used the same range for the x- and y-axis). It is clear that, even though the resolution used by WVHS03 during the scan of the parameter space in $P_{\rm orb}$ is higher than the one adopted here, the magnitude and trend of the timescales due to dynamic tides both in and out of resonance are in good agreement (and, as pointed out by WVHS03, also in agreement with what previous investigations have found, e.g. \citealt{SavonijePapaloizou1983}). For the results presented by WVHS03, the resonantly excited eigenmodes in the range $P_{\rm orb}\,=\,$2~-5~days are $g-$modes of radial order n from 1 to 12. In Table~\ref{Tab:MESA_ZAMS_5Msun_Zsun_Xc0.7_log23_eigenfreq_nonAd} we list the real part of the non-adiabatic eigenfrequencies for the modes $g_{\rm 1}-g_{\rm 12}$ of the stellar model adopted in this work. As our stellar model differs from the one used by WVHS03, a direct comparison between their eigenfrequencies (see Table~1 of WVHS03) and the one calculated here is not possible. However, 
making WVHS03's eigenfrequencies dimensionless, $\omega_{\rm R}$ for the $g_{\rm 1}$- and $g_{\rm 12}$-modes calculated by WVHS03 are 2.20672 and 0.36723, respectively. This suggests that the resonantly excited modes in our stellar model are in the range between $g_{\rm 1}$ to $g_{\rm 11}$, in agreement between the two studies.
\begin{deluxetable}{c|c|c|c}
\tablecolumns{4}
\tablewidth{0pc}
\tablecaption{Real component of the non-adiabatic $l\,=\,2$ eigenfrequencies for the 5$M_\odot$ MS described in \S~\ref{Dynamic Tides Timescales in an Eccentric Binary Hosting a 5Msun MS star and a Neutron Star}. }
\tablehead{
\colhead{mode} & \colhead{$\omega_{\rm R}$}&\colhead{mode} & \colhead{$\omega_{\rm R}$}}
\startdata
$g_{\rm 1}$&  2.04543   &$g_{\rm 7}$ & 5.35948~$\times~10^{-1}$\\
$g_{\rm 2}$ & 1.41657   &$g_{\rm 8}$ & 4.78377~$\times~10^{-1}$\\
$g_{\rm 3}$ & 1.07685  &$g_{\rm 9}$ & 4.32999~$\times~10^{-1}$\\
$g_{\rm 4}$ & 8.61857~$\times~10^{-1}$ & $g_{\rm 10}$ & 3.95316~$\times~10^{-1}$\\
$g_{\rm 5}$ & 7.16287~$\times~10^{-1}$&$g_{\rm 11}$ & 3.63316~$\times~10^{-1}$\\
$g_{\rm 6}$ & 6.12283~$\times~10^{-1}$&$g_{\rm 12}$ & 3.36272~$\times~10^{-1}$
\enddata
\label{Tab:MESA_ZAMS_5Msun_Zsun_Xc0.7_log23_eigenfreq_nonAd}
\end{deluxetable}
\subsubsection{Dynamic Tides Timescales in a Binary Hosting a 1.5$\,M_{\odot}$ MS star and a Hot Jupiter}\label{Dynamic Tides Timescales in an Eccentric Binary Hosting a 1.5Msun MS star and a hot Jupiter}

As of Spring 2013, $\simeq\,$850 planets have been confirmed and $\simeq\,$2,700 new candidates have been provided by NASA's \emph{Kepler} satellite \citep{BatalhaKeplerTeam2012}. 
Among the confirmed exoplanets, $\simeq\,$80 have a mass $\gtrsim\,M_{\rm J}$ and an orbital period $P_{\rm orb}\,\lesssim\,5\,$d. This number increases with a less stringent mass and period constraints and, potentially, many more hot Jupiters exist among the planet candidates.
The so-called ``very hot Jupiters'' have $P_{\rm orb}\,\lesssim\,$1~d, e.g. \citep{HellierACGHMQSTWWBEHILMPPPSUW2009Natur}. 

In these planetary systems, the tidal forcing frequencies induced by the giant planet in the host star are resonant with high order $g$-modes. Our goal here is to demonstrate that {\tt CAFein} can handle such high order modes. As {\tt CAFein}'s current state neglect the effect of convection, we consider a star with an envelope in radiative equilibrium.  

\begin{figure} [!h]
  \epsscale{1.1}
\plotone{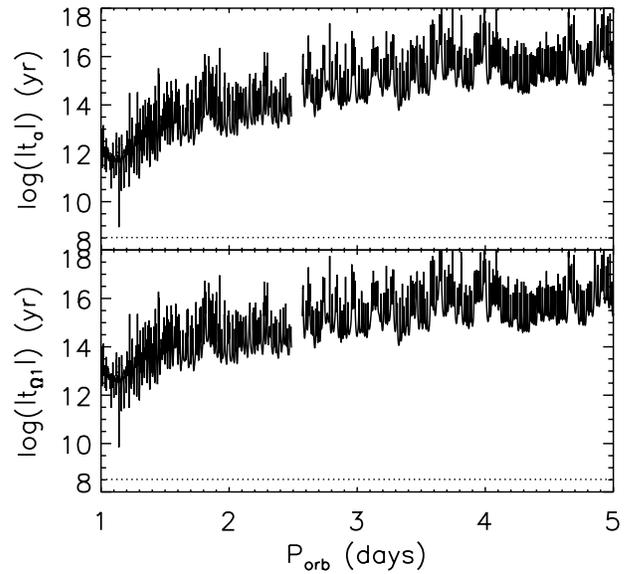}
\caption{Orbital and spin evolution timescales due to dynamic tides for a binary hosting a 1.5$\,M_{\odot}$ MS star and a hot Jupiter in a circular orbit as a function of the orbital period ($P_{\rm orb}$). The timescales are as in Fig.~\ref{fig:Comparison_WillemEtAl2003_Valsecchi_kUpTo9}. The horizontal dotted line represents the logarithm of the star's age. The gap in the timescales at $\simeq~2.5\,$days is due to the resolution adopted during the calculation.}
\label{fig:hotJupiter_star}
\end{figure}

We use MESA to create a stellar model of a 1.5$\,M_{\odot}$ MS star at solar metallicity and increase the number of mesh points as described in \S~\ref{Testing CAFein on a Zero Age Main Sequence Star: the Non Adiabatic Case}.
The radius of our model is $R_{\rm 1}\,\simeq\,1.49\,R_{\odot}$ and its age is $\tau_{*}$=3.3$\times 10^{8}\,$yr. We take the companion to be a point mass of 1 $M_{\rm J}$, we set the star's spin to be 50\% of the companion's orbital angular velocity, and we take the orbit to be circular. 
We consider the dominant term in the expansion of the tide-generating potential and fix $(l, m, k)~=~(2, -2, 2)$. We then calculate the orbital and spin evolution timescales for orbital periods ranging from 1 to 5\,days. The results are summarized in Fig.~\ref{fig:hotJupiter_star}. 
The orbital and spin periods considered correspond to a range in tidal forcing frequencies between $\omega_{T}\simeq 0.04-0.18$, which for the stellar model adopted here spans modes between $\simeq g_{50} - g_{200}$.  

Even though our goal here was to demonstrate only that {\tt CAFein} can handle the high-order modes involved when dynamic tides in a star hosting a hot Jupiter are considered, we note that we currently neglect the effect of rotation on the eigenfrequency spectrum. Depending on its magnitude, rotation is expected to affect the magnitude of the timescales computed, as it enriches the eigenfrequencies spectrum leading to more resonances \citep{WitteSavonije1999b,WitteSavonije1999}.

\section{Summary, Discussion, and Conclusions}\label{CAFein: Discussion and Conclusions}

Here we have presented {\tt CAFein}, a new computational tool for calculating non-adiabatic stellar oscillations in isolated stars and tidally excited stellar oscillations in close binaries, particularly in the dynamic tides regime, where the driving frequencies are comparable to the stellar eigenfrequencies. Tides are considered as a small perturbation applied on a spherically symmetric star in hydrostatic equilibrium and the linear approximation is adopted. {\tt CAFein} is based on the so-called Riccati method, a numerical algorithm that has been extensively and successfully applied to a variety of stellar pulsators and which does not suffer from the major drawbacks of commonly-used shooting and relaxation schemes. 
Even though the Riccati method is formally a shooting method, it relies on transforming the linear first-order boundary eigenvalue problem describing stellar oscillations into a numerically stable, non-linear initial value problem. This initial-value problem is then solved using a shooting method, where the eigenfrequency is the only shooting parameter to be iterated. 

The inclusion of the tide-generating potential in the stellar pulsation equations formally does not change the system of equations that have to be integrated, and thus the applicability of the Riccati method. However, it renders the BCs at the star's surface non-homogeneous. We made the Riccati method viable for solving the tidally excited stellar pulsation problem by introducing two new variables and corresponding differential equations to make the BCs homogeneous.

We tested {\tt CAFein} as a pure stellar pulsation code for two different applications. In the adiabatic regime, we first calculated the eigenfrequencies of a polytrope and verified that the results are not significantly affected if some of the relevant parameters entering the Riccati method are varied, thus demonstrating {\tt CAFein}'s numerical stability. Next, we compared the computed eigenfrequencies with previously published results which relied both on the Riccati method (TL04) and on other commonly-used shooting and relaxation techniques (CDM94). The comparison yielded very good agreement and the orthogonality of low order eigenfunctions was also successfully verified.
In the non-adiabatic regime, we considered a stellar model in the 
$\beta$ Cephei/SPB instability strip of the HR diagram and successfully recovered the unstable modes in the part of the parameter space examined.  

We showed that the extension of the Riccati method to treat tidally excited stellar pulsations works as expected by successfully reproducing the work presented by PS90 and showing that a dynamical tide can be approximated as the sum of the equilibrium tide and another part reflecting the oscillatory properties of the star itself. Furthermore, we successfully reproduced the magnitude and trend of the orbital and spin evolution timescales due to dynamic tides both in and out of resonance presented by WVHS03. Here we stress the fact that WVHS03 did not solve numerically the fully non-adiabatic problem but found the perturbed stellar quantities via semi-analytical solutions. Finally, we applied {\tt CAFein} to a 1.5$\, M_{\odot}$ Main Sequence star hosting a hot Jupiter and showed that this code can handle the high-order modes involved in these binaries.

In this paper we have explored {\tt CAFein}'s performance in the dynamic tides regime, where the tidal forcing frequencies are close to the star's eigenfrequencies and the latter \emph{can} be resonantly excited by tides. This code could be used to explore cases that are closer to the quasi-static tides limit, in which the tidal forcing frequencies are much smaller compared to the inverse of the WD's dynamical time scale (almost synchronized components or long orbital and rotational periods).  However, as noted by \cite{SavonijePapaloizou1983}, it is  numerically challenging to calculate low-frequency tides by
integrating the full set of tidal oscillation equations because of the short-wavelength components entering the tidal response. For this reasons, the system of equations is usually reduced using perturbation theory (e.g. \citealt{Smeyers1997,WVHS2003,WDK2010}).

Even though the physics included in {\tt CAFein} makes it suitable for investigations of stars with envelopes that are mostly radiative we intend to use this novel code to investigate a variety of binaries and stars. To this purpose, we are currently upgrading it to account for the effect of rotation on the eigenmode spectrum in the so-called \textit{traditional approximation} \citep{UnnoEtAl1989} and the effect of turbulent friction acting on the equilibrium tide \citep{TerquemPN1998, SavonijeWitte2002, WDK2010}.

\acknowledgements
 We are grateful to A. Barker, Y. Lithwick, Lars Bildsten, Bill Paxton, and the MESA community for useful discussions during the
 development of {\tt CAFein}. Simulations were performed on the
 computing cluster {\tt Fugu} available to the Theoretical
 Astrophysics group at Northwestern and partially funded by NSF grant
 PHY--0619274 to VK. This work was supported by NASA Awards NNX09AJ56G to V.K. and NNX12AI86G to F.A.R.

\begin{acknowledgements}
\end{acknowledgements}
\newpage
\bibliographystyle{apj}

\begin{thebibliography}{81}
\expandafter\ifx\csname natexlab\endcsname\relax\def\natexlab#1{#1}\fi

\bibitem[{{Amaro-Seoane} {et~al.}(2012){Amaro-Seoane}, {Aoudia}, {Babak},
  {Bin{\'e}truy}, {Berti}, {Boh{\'e}}, {Caprini}, {Colpi}, {Cornish},
  {Danzmann}, {Dufaux}, {Gair}, {Jennrich}, {Jetzer}, {Klein}, {Lang}, {Lobo},
  {Littenberg}, {McWilliams}, {Nelemans}, {Petiteau}, {Porter}, {Schutz},
  {Sesana}, {Stebbins}, {Sumner}, {Vallisneri}, {Vitale}, {Volonteri}, \&
  {Ward}}]{NGOeLISA2012}
{Amaro-Seoane}, P., {et~al.} 2012, Classical and Quantum Gravity, 29, 124016

\bibitem[{{Barker} \& {Ogilvie}(2010)}]{BarkerOgilvie2010}
{Barker}, A.~J., \& {Ogilvie}, G.~I. 2010, \mnras, 404, 1849

\bibitem[{{Batalha} \& {Kepler Team}(2012)}]{BatalhaKeplerTeam2012}
{Batalha}, N.~M., \& {Kepler Team}. 2012, in American Astronomical Society
  Meeting Abstracts, Vol. 220, American Astronomical Society Meeting Abstracts
  $\#$220, 306.01

\bibitem[{{Brassard} {et~al.}(1991){Brassard}, {Fontaine}, {Wesemael},
  {Kawaler}, \& {Tassoul}}]{BrassardEtAl1991}
{Brassard}, P., {Fontaine}, G., {Wesemael}, F., {Kawaler}, S.~D., \& {Tassoul},
  M. 1991, \apj, 367, 601

\bibitem[{{Brown} {et~al.}(2010){Brown}, {Kilic}, {Allende Prieto}, \&
  {Kenyon}}]{BrownEtAl2010}
{Brown}, W.~R., {Kilic}, M., {Allende Prieto}, C., \& {Kenyon}, S.~J. 2010,
  \apj, 723, 1072

\bibitem[{{Brown} {et~al.}(2012){Brown}, {Kilic}, {Allende Prieto}, \&
  {Kenyon}}]{BrownEtAl2012}
---. 2012, \apj, 744, 142

\bibitem[{{Brown} {et~al.}(2011){Brown}, {Kilic}, {Hermes}, {Allende Prieto},
  {Kenyon}, \& {Winget}}]{BrownEtAl2011}
{Brown}, W.~R., {Kilic}, M., {Hermes}, J.~J., {Allende Prieto}, C., {Kenyon},
  S.~J., \& {Winget}, D.~E. 2011, \apjl, 737, L23+

\bibitem[{{Burkart} {et~al.}(2012{\natexlab{a}}){Burkart}, {Quataert}, {Arras},
  \& {Weinberg}}]{BurkartEtAl2012HeWD}
{Burkart}, J., {Quataert}, E., {Arras}, P., \& {Weinberg}, N.~N.
  2012{\natexlab{a}}, ArXiv e-prints

\bibitem[{{Burkart} {et~al.}(2012{\natexlab{b}}){Burkart}, {Quataert}, {Arras},
  \& {Weinberg}}]{BurkartEtAl2012}
---. 2012{\natexlab{b}}, \mnras, 421, 983

\bibitem[{{Cantiello} {et~al.}(2009){Cantiello}, {Langer}, {Brott}, {de Koter},
  {Shore}, {Vink}, {Voegler}, {Lennon}, \& {Yoon}}]{CantielloLBdKSVVLY2009}
{Cantiello}, M., {et~al.} 2009, \aap, 499, 279

\bibitem[{{Christensen-Dalsgaard} \& {Mullan}(1994)}]{CDM1994}
{Christensen-Dalsgaard}, J., \& {Mullan}, D.~J. 1994, \mnras, 270, 921

\bibitem[{{Cox} {et~al.}(1992){Cox}, {Morgan}, {Rogers}, \&
  {Iglesias}}]{CMRI1992}
{Cox}, A.~N., {Morgan}, S.~M., {Rogers}, F.~J., \& {Iglesias}, C.~A. 1992,
  \apj, 393, 272

\bibitem[{{Danzmann} \& {the LISA study team}(1996)}]{Danzmann1996}
{Danzmann}, K., \& {the LISA study team}. 1996, Classical and Quantum Gravity,
  13, 247

\bibitem[{{Davey}(1977)}]{Davey1977}
{Davey}, A. 1977, Journal of Computational Physics, 24, 331

\bibitem[{{Dziembowski} {et~al.}(1993){Dziembowski}, {Moskalik}, \&
  {Pamyatnykh}}]{DziembowskiMP1993}
{Dziembowski}, W.~A., {Moskalik}, P., \& {Pamyatnykh}, A.~A. 1993, \mnras, 265,
  588

\bibitem[{{Fuller} \& {Lai}(2011)}]{FullerLai2011a}
{Fuller}, J., \& {Lai}, D. 2011, \mnras, 412, 1331

\bibitem[{{Fuller} \& {Lai}(2012{\natexlab{a}})}]{FullerLai2012HeWD}
---. 2012{\natexlab{a}}, ArXiv e-prints

\bibitem[{{Fuller} \& {Lai}(2012{\natexlab{b}})}]{FullerLai2012Dynamic}
---. 2012{\natexlab{b}}, \mnras, 421, 426

\bibitem[{{Fuller} \& {Lai}(2012{\natexlab{c}})}]{FullerLai2012KOI54}
---. 2012{\natexlab{c}}, \mnras, 420, 3126

\bibitem[{{Gautschy} \& {Glatzel}(1990{\natexlab{a}})}]{GautschyGlatzel1990}
{Gautschy}, A., \& {Glatzel}, W. 1990{\natexlab{a}}, \mnras, 245, 154

\bibitem[{{Gautschy} \& {Glatzel}(1990{\natexlab{b}})}]{GautschyGlatzel1990b}
---. 1990{\natexlab{b}}, \mnras, 245, 597

\bibitem[{{Gautschy} \& {Glatzel}(1991)}]{GautschyGlatzel1991}
---. 1991, \mnras, 253, 509

\bibitem[{{Gautschy} \& {L{\"o}ffler}(1996)}]{GautschyLoeffler1996}
{Gautschy}, A., \& {L{\"o}ffler}, W. 1996, Delta Scuti Star Newsletter, 10, 13

\bibitem[{{Gautschy} {et~al.}(1996){Gautschy}, {Ludwig}, \&
  {Freytag}}]{GLF1996}
{Gautschy}, A., {Ludwig}, H.-G., \& {Freytag}, B. 1996, \aap, 311, 493

\bibitem[{{Gautschy} \& {Saio}(1995)}]{GautschySaio1995}
{Gautschy}, A., \& {Saio}, H. 1995, \araa, 33, 75

\bibitem[{{Gautschy} \& {Saio}(1996)}]{GautschySaio1996}
---. 1996, \araa, 34, 551

\bibitem[{{Giuricin} {et~al.}(1984){Giuricin}, {Mardirossian}, \&
  {Mezzetti}}]{GiuricinEtAl1984}
{Giuricin}, G., {Mardirossian}, F., \& {Mezzetti}, M. 1984, \aap, 134, 365

\bibitem[{{Glatzel} \& {Gautschy}(1992)}]{GlatzelGautschy1992}
{Glatzel}, W., \& {Gautschy}, A. 1992, \mnras, 256, 209

\bibitem[{{Glatzel} \& {Kiriakidis}(1993)}]{GlatzelKiriakidis1993}
{Glatzel}, W., \& {Kiriakidis}, M. 1993, \mnras, 263, 375

\bibitem[{{Goodman} \& {Dickson}(1998)}]{GoodmanDickson1998}
{Goodman}, J., \& {Dickson}, E.~S. 1998, \apj, 507, 938

\bibitem[{{Hellier} {et~al.}(2009){Hellier}, {Anderson}, {Collier Cameron},
  {Gillon}, {Hebb}, {Maxted}, {Queloz}, {Smalley}, {Triaud}, {West}, {Wilson},
  {Bentley}, {Enoch}, {Horne}, {Irwin}, {Lister}, {Mayor}, {Parley}, {Pepe},
  {Pollacco}, {Segransan}, {Udry}, \&
  {Wheatley}}]{HellierACGHMQSTWWBEHILMPPPSUW2009Natur}
{Hellier}, C., {et~al.} 2009, \nat, 460, 1098

\bibitem[{{Henyey} {et~al.}(1964){Henyey}, {Forbes}, \& {Gould}}]{HFG1964}
{Henyey}, L.~G., {Forbes}, J.~E., \& {Gould}, N.~L. 1964, \apj, 139, 306

\bibitem[{{Hermes} {et~al.}(2012){Hermes}, {Kilic}, {Brown}, {Winget}, {Allende
  Prieto}, {Gianninas}, {Mukadam}, {Cabrera-Lavers}, \&
  {Kenyon}}]{HermesEtAl2012}
{Hermes}, J.~J., {et~al.} 2012, ArXiv e-prints

\bibitem[{{Hut}(1981)}]{Hut1981}
{Hut}, P. 1981, \aap, 99, 126

\bibitem[{{Iglesias} {et~al.}(1992){Iglesias}, {Rogers}, \& {Wilson}}]{IRW1992}
{Iglesias}, C.~A., {Rogers}, F.~J., \& {Wilson}, B.~G. 1992, \apj, 397, 717

\bibitem[{{Kilic} {et~al.}(2011){Kilic}, {Brown}, {Allende Prieto},
  {Ag{\"u}eros}, {Heinke}, \& {Kenyon}}]{KilicEtAl2011}
{Kilic}, M., {Brown}, W.~R., {Allende Prieto}, C., {Ag{\"u}eros}, M.~A.,
  {Heinke}, C., \& {Kenyon}, S.~J. 2011, \apj, 727, 3

\bibitem[{{Kilic} {et~al.}(2012){Kilic}, {Brown}, {Allende Prieto}, {Kenyon},
  {Heinke}, {Ag{\"u}eros}, \& {Kleinman}}]{KilicEtAl2012}
{Kilic}, M., {Brown}, W.~R., {Allende Prieto}, C., {Kenyon}, S.~J., {Heinke},
  C.~O., {Ag{\"u}eros}, M.~A., \& {Kleinman}, S.~J. 2012, \apj, 751, 141

\bibitem[{{Kilic} {et~al.}(2010){Kilic}, {Brown}, {Allende Prieto}, {Kenyon},
  \& {Panei}}]{KilicEtAl2010}
{Kilic}, M., {Brown}, W.~R., {Allende Prieto}, C., {Kenyon}, S.~J., \& {Panei},
  J.~A. 2010, \apj, 716, 122

\bibitem[{{Kiriakidis} {et~al.}(1992){Kiriakidis}, {El Eid}, \&
  {Glatzel}}]{KEEG1992}
{Kiriakidis}, M., {El Eid}, M.~F., \& {Glatzel}, W. 1992, \mnras, 255, 1P

\bibitem[{{Liu} {et~al.}(2010){Liu}, {Han}, {Zhang}, \& {Zhang}}]{LiuEtAl2010b}
{Liu}, J., {Han}, Z., {Zhang}, F., \& {Zhang}, Y. 2010, \apj, 719, 1546

\bibitem[{{L{\"o}ffler}(2000)}]{Loffler2000}
{L{\"o}ffler}, W. 2000, in Astronomical Society of the Pacific Conference
  Series, Vol. 203, IAU Colloq. 176: The Impact of Large-Scale Surveys on
  Pulsating Star Research, ed. L.~{Szabados} \& D.~{Kurtz}, 447--448

\bibitem[{{Mazeh} {et~al.}(2006){Mazeh}, {Tamuz}, \& {North}}]{MazehTN2006}
{Mazeh}, T., {Tamuz}, O., \& {North}, P. 2006, \mnras, 367, 1531

\bibitem[{{Moskalik} \& {Dziembowski}(1992)}]{MoskalikDziembowski1992}
{Moskalik}, P., \& {Dziembowski}, W.~A. 1992, \aap, 256, L5

\bibitem[{{Nelemans} {et~al.}(2001{\natexlab{a}}){Nelemans}, {Portegies Zwart},
  {Verbunt}, \& {Yungelson}}]{Nelemans2001b}
{Nelemans}, G., {Portegies Zwart}, S.~F., {Verbunt}, F., \& {Yungelson}, L.~R.
  2001{\natexlab{a}}, \aap, 368, 939

\bibitem[{{Nelemans} {et~al.}(2004){Nelemans}, {Yungelson}, \& {Portegies
  Zwart}}]{NelemansEtAl2004}
{Nelemans}, G., {Yungelson}, L.~R., \& {Portegies Zwart}, S.~F. 2004, \mnras,
  349, 181

\bibitem[{{Nelemans} {et~al.}(2001{\natexlab{b}}){Nelemans}, {Yungelson},
  {Portegies Zwart}, \& {Verbunt}}]{Nelemans2001a}
{Nelemans}, G., {Yungelson}, L.~R., {Portegies Zwart}, S.~F., \& {Verbunt}, F.
  2001{\natexlab{b}}, \aap, 365, 491

\bibitem[{{North} \& {Zahn}(2003)}]{NorthZahn2003}
{North}, P., \& {Zahn}, J.-P. 2003, \aap, 405, 677

\bibitem[{{Paxton} {et~al.}(2011){Paxton}, {Bildsten}, {Dotter}, {Herwig},
  {Lesaffre}, \& {Timmes}}]{PBDHLT2011}
{Paxton}, B., {Bildsten}, L., {Dotter}, A., {Herwig}, F., {Lesaffre}, P., \&
  {Timmes}, F. 2011, \apjs, 192, 3

\bibitem[{{Polfliet} \& {Smeyers}(1990)}]{PolflietSmeyers1990}
{Polfliet}, R., \& {Smeyers}, P. 1990, \aap, 237, 110

\bibitem[{{Rathore} {et~al.}(2005){Rathore}, {Blandford}, \&
  {Broderick}}]{RathoreEtAl2005}
{Rathore}, Y., {Blandford}, R.~D., \& {Broderick}, A.~E. 2005, \mnras, 357, 834

\bibitem[{{Rogers} \& {Iglesias}(1992)}]{RogersIglesias1992}
{Rogers}, F.~J., \& {Iglesias}, C.~A. 1992, \apjs, 79, 507

\bibitem[{{Ruiter} {et~al.}(2010){Ruiter}, {Belczynski}, {Benacquista},
  {Larson}, \& {Williams}}]{RuiterEtAl2010}
{Ruiter}, A.~J., {Belczynski}, K., {Benacquista}, M., {Larson}, S.~L., \&
  {Williams}, G. 2010, \apj, 717, 1006

\bibitem[{{Ruymaekers}(1992)}]{Ruymaekers1992}
{Ruymaekers}, E. 1992, \aap, 259, 349

\bibitem[{{Saio} \& {Cox}(1980)}]{SaioCox1980}
{Saio}, H., \& {Cox}, J.~P. 1980, \apj, 236, 549

\bibitem[{{Savonije} \& {Papaloizou}(1983)}]{SavonijePapaloizou1983}
{Savonije}, G.~J., \& {Papaloizou}, J.~C.~B. 1983, \mnras, 203, 581

\bibitem[{{Savonije} \& {Witte}(2002)}]{SavonijeWitte2002}
{Savonije}, G.~J., \& {Witte}, M.~G. 2002, \aap, 386, 211

\bibitem[{{Schenker} \& {Gautschy}(1998)}]{SchenkerGautschy1998}
{Schenker}, K., \& {Gautschy}, A. 1998, in Astronomical Society of the Pacific
  Conference Series, Vol. 135, A Half Century of Stellar Pulsation
  Interpretation, ed. P.~A. {Bradley} \& J.~A. {Guzik}, 116

\bibitem[{{Scott}(1973)}]{Scott1973}
{Scott}, M.~R. 1973, Journal of Computational Physics, 12, 334

\bibitem[{{Seaton} {et~al.}(1994){Seaton}, {Yan}, {Mihalas}, \&
  {Pradhan}}]{SYMP1994}
{Seaton}, M.~J., {Yan}, Y., {Mihalas}, D., \& {Pradhan}, A.~K. 1994, \mnras,
  266, 805

\bibitem[{{Sloan}(1977)}]{Sloan1977}
{Sloan}, D.~M. 1977, Journal of Computational Physics, 24, 320

\bibitem[{{Smeyers}(1997)}]{Smeyers1997}
{Smeyers}, P. 1997, \aap, 318, 140

\bibitem[{{Smeyers} \& {Willems}(1998)}]{SmeyersWillems1998}
{Smeyers}, P., \& {Willems}, B. 1998, \aap, 336, 539

\bibitem[{{Smeyers} {et~al.}(1998){Smeyers}, {Willems}, \& {Van
  Hoolst}}]{SWVH1998}
{Smeyers}, P., {Willems}, B., \& {Van Hoolst}, T. 1998, \aap, 335, 622

\bibitem[{{Steffen}(1990)}]{Steffen1990}
{Steffen}, M. 1990, \aap, 239, 443

\bibitem[{{Stothers} \& {Chin}(1993)}]{StothersChin1993}
{Stothers}, R.~B., \& {Chin}, C.-W. 1993, \apjl, 408, L85

\bibitem[{{Takata} \& {L{\"o}ffler}(2004)}]{TakataLoffler2004}
{Takata}, M., \& {L{\"o}ffler}, W. 2004, \pasj, 56, 645

\bibitem[{{Terquem} {et~al.}(1998){Terquem}, {Papaloizou}, {Nelson}, \&
  {Lin}}]{TerquemPN1998}
{Terquem}, C., {Papaloizou}, J.~C.~B., {Nelson}, R.~P., \& {Lin}, D.~N.~C.
  1998, \apj, 502, 788

\bibitem[{{Thompson} {et~al.}(2012){Thompson}, {Everett}, {Mullally},
  {Barclay}, {Howell}, {Still}, {Rowe}, {Christiansen}, {Kurtz}, {Hambleton},
  {Twicken}, {Ibrahim}, \& {Clarke}}]{TEMBHSRCKHTIC2012}
{Thompson}, S.~E., {et~al.} 2012, \apj, 753, 86

\bibitem[{{Unno} {et~al.}(1989){Unno}, {Osaki}, {Ando}, {Saio}, \&
  {Shibahashi}}]{UnnoEtAl1989}
{Unno}, W., {Osaki}, Y., {Ando}, H., {Saio}, H., \& {Shibahashi}, H. 1989,
  {Nonradial oscillations of stars}, ed. {Unno, W., Osaki, Y., Ando, H., Saio,
  H., \& Shibahashi, H.}

\bibitem[{{Valsecchi} {et~al.}(2012){Valsecchi}, {Farr}, {Willems}, \&
  {Kalogera}}]{ValsecchiFWKJ06512012}
{Valsecchi}, F., {Farr}, W.~M., {Willems}, B., \& {Kalogera}, V. 2012, ArXiv
  e-prints

\bibitem[{{Welsh} {et~al.}(2011){Welsh}, {Orosz}, {Aerts}, {Brown},
  {Brugamyer}, {Cochran}, {Gilliland}, {Guzik}, {Kurtz}, {Latham}, {Marcy},
  {Quinn}, {Zima}, {Allen}, {Batalha}, {Bryson}, {Buchhave}, {Caldwell},
  {Gautier}, {Howell}, {Kinemuchi}, {Ibrahim}, {Isaacson}, {Jenkins}, {Prsa},
  {Still}, {Street}, {Wohler}, {Koch}, \& {Borucki}}]{WelshEtAl2011}
{Welsh}, W.~F., {et~al.} 2011, \apjs, 197, 4

\bibitem[{{Willems}(2003)}]{Willems2003}
{Willems}, B. 2003, \mnras, 346, 968

\bibitem[{{Willems} {et~al.}(2010){Willems}, {Deloye}, \& {Kalogera}}]{WDK2010}
{Willems}, B., {Deloye}, C.~J., \& {Kalogera}, V. 2010, \apj, 713, 239

\bibitem[{{Willems} {et~al.}(2003){Willems}, {van Hoolst}, \&
  {Smeyers}}]{WVHS2003}
{Willems}, B., {van Hoolst}, T., \& {Smeyers}, P. 2003, \aap, 397, 973

\bibitem[{{Witte} \& {Savonije}(1999{\natexlab{a}})}]{WitteSavonije1999b}
{Witte}, M.~G., \& {Savonije}, G.~J. 1999{\natexlab{a}}, \aap, 341, 842

\bibitem[{{Witte} \& {Savonije}(1999{\natexlab{b}})}]{WitteSavonije1999}
---. 1999{\natexlab{b}}, \aap, 350, 129

\bibitem[{{Witte} \& {Savonije}(2001)}]{WitteSavonije2002TwoRitatingMSstars}
---. 2001, \aap, 366, 840

\bibitem[{{Zahn}(1975)}]{Zahn1975}
{Zahn}, J.-P. 1975, \aap, 41, 329

\bibitem[{{Zahn}(1977)}]{Zahn1977}
---. 1977, \aap, 57, 383

\bibitem[{{Zahn}(2008)}]{Zahn2008}
{Zahn}, J.-P. 2008, in EAS Publications Series, Vol.~29, EAS Publications
  Series, ed. M.-J. {Goupil} \& J.-P. {Zahn}, 67--90

\bibitem[{{Zahn}(1978)}]{Zahn1978}
{Zahn}, J.-R. 1978, \aap, 67, 162

\end{thebibliography}

\end{document}